\documentclass[apj]{emulateapj}
\usepackage[colorlinks,linkcolor=blue]{hyperref}
\usepackage{bm}
\usepackage{graphicx}
\usepackage{epsf}
\usepackage{graphics}
\usepackage{amsmath}

\def\beq{\begin{equation}}
\def\eeq{\end{equation}}
\def\bey{\begin{eqnarray}}
\def\eey{\end{eqnarray}}

\def\mpc{\, h^{-1}{\rm {Mpc}}}
\def\mpchi{\, h{\rm {Mpc}}^{-1}}

\def\kms{\,{\rm {km\, s^{-1}}}}

\def\msun{\, h^{-1}{\rm M_\odot}}

\def\gs{\mathrel{\raise1.16pt\hbox{$>$}\kern-7.0pt
\lower3.06pt\hbox{{$\scriptstyle \sim$}}}}
\def\ls{\mathrel{\raise1.16pt\hbox{$<$}\kern-7.0pt
\lower3.06pt\hbox{{$\scriptstyle \sim$}}}}
\def\gtsima{$\; \buildrel > \over \sim \;$}
\def\ltsima{$\; \buildrel < \over \sim \;$}
\def\prosima{$\; \buildrel \propto \over \sim \;$}
\def\gsim{\lower.5ex\hbox{\gtsima}}
\def\lsim{\lower.5ex\hbox{\ltsima}}
\def\simgt{\lower.5ex\hbox{\gtsima}}
\def\simlt{\lower.5ex\hbox{\ltsima}}
\def\simpr{\lower.5ex\hbox{\prosima}}

\shorttitle{Exploring the Local Universe with reConstructed Initial Density field}
\shortauthors{Wang H.Y. et al.}

\begin{document}

\title {ELUCID - Exploring the Local Universe with reConstructed Initial Density field
III: Constrained Simulation in the SDSS Volume}
\author{Huiyuan Wang\altaffilmark{1}, H.J. Mo\altaffilmark{2,3}, Xiaohu Yang\altaffilmark{4,5}, Youcai Zhang\altaffilmark{6}, JingJing Shi\altaffilmark{1,7}, Y. P. Jing\altaffilmark{4,5}, Chengze Liu\altaffilmark{4}, Shijie Li\altaffilmark{4}, Xi Kang\altaffilmark{8} and Yang Gao\altaffilmark{9}}

\altaffiltext{1}{Key Laboratory for Research in Galaxies and Cosmology, Department of Astronomy, University of Science and
Technology of China, Hefei, Anhui 230026, China; whywang@mail.ustc.edu.cn}

\altaffiltext{2}{Department of Astronomy, University of Massachusetts, Amherst MA 01003-9305, USA}

\altaffiltext{3}{Astronomy Department and Center for Astrophysics, Tsinghua University, Beijing 10084, China}

\altaffiltext{4}{Center for Astronomy and Astrophysics, Shanghai Jiao Tong University, Shanghai 200240, China}

\altaffiltext{5}{IFSA Collaborative Innovation Center, Shanghai Jiao Tong University, Shanghai 200240, China}

\altaffiltext{6}{Key Laboratory for Research in Galaxies and Cosmology, Shanghai Astronomical Observatory, Nandan Road 80, Shanghai 200030, China}

\altaffiltext{7}{SISSA, Via Bonomea 265, I-34136 Trieste, Italy}

\altaffiltext{8}{Purple Mountain Observatory, the Partner Group of MPI f$\ddot{u}$r Astronomie, 2 West Beijing Road, Nanjing 210008, China}

\altaffiltext{9}{The Computer Network Information Center of the Chinese Academy of Sciences, Beijing 100190, China}

\begin{abstract}
A method we developed recently for the reconstruction of the initial density field
in the nearby Universe is applied to the Sloan Digital Sky Survey Data Release 7.
A high-resolution N-body constrained simulation (CS) of the reconstructed initial condition, with $3072^3$ particles evolved in a
$500h^{-1}{\rm Mpc}$ box, is carried out
and analyzed in terms of the statistical properties of the final density field and its
relation with the distribution of SDSS galaxies. We find that the statistical properties
of the cosmic web and the halo populations are accurately reproduced in the CS.
The galaxy density field is strongly correlated with the CS density field, with a bias that depend
on both galaxy luminosity and color. Our further investigations show that the CS provides
robust quantities describing the environments within which the observed galaxies
and galaxy systems reside. Cosmic variance is greatly reduced in the CS so that
the statistical uncertainties can be controlled effectively even for samples of
small volumes.
\end{abstract}

\keywords{dark matter - large-scale structure of the universe -
galaxies: halos - methods: statistical}

\section{Introduction}
\label{sec_intro}

One of the major aims of modern astrophysics is to understand how galaxies and galaxy
systems form and evolve in the cosmic density field. In the current standard cosmogony, galaxies
are assumed to form via radiative cooling and collapse of baryonic gas within the gravitational
potentials of the dark matter structures (e.g. White \& Rees 1978).  Therefore exploring the
relationships and interactions between baryonic gas, galaxies and dark matter is not only essential
for understanding galaxy formation (e.g. Mo, van den Bosch \& White 2010), but also important for
using galaxies as tracers of the large-scale structure of the universe to test models for cosmology and dark matter particles
(e.g., Peebles 1980;  Peebles \& Nusser 2010; Frenk \& White 2012).

Many observational programs have been carried out to study the properties of galaxies and
the baryonic gas, such as the intergalactic medium (IGM), circum-galactic medium (CGM)
and  interstellar medium (ISM). Galaxy properties are observed through multi-band galaxy
surveys, such as the Sloan Digital Sky Survey (SDSS; York et al. 2000) and the Two Micron
All Sky Survey (Skrutskie et al. 2006). The gas components can be detected via various ways,
such as X-ray observations (e.g. Rosati et al. 2002), quasar absorption line systems
(e.g. Danforth \& Shull 2008), the 21 cm emission of neutral hydrogen gas (e.g., Koribalski et al. 2004),
and millimeter/sub-millimeter emissions of molecular gas (e.g. Young et al. 1995). These observations
provide unprecedented data base to study the processes that shape the eco-systems of galaxies
we observe.

Theoretically, great progress has been made in understanding the structures in cold dark matter
distribution (e.g. Zel'dovich 1970; Davis et al. 1985; Springel et al. 2005). The key concept in the buildup
of the cosmic structure is the formation of dark matter halos, which are the hosts of
galaxies. With the help of N-body simulations, the properties of the halo population, such as
the spatial clustering, mass function, assembly history, and internal structure, are well understood
(e.g. Mo et al. 2010 for a review). Within the current $\Lambda$CDM paradigm, methods, such as
hydrodynamical simulations (Katz 1992; Navarro \& White 1993; Vogelsberger et al. 2014; Schaye  et al. 2015),
semi-analytic model (White \& Frenk 1991; Croton et al. 2006; Lu et al. 2011),
and empirical models  (Jing et al. 1998; Peacock \& Smith 2000; Yang et al. 2003;
Kravtsov et al. 2004; Zheng et al. 2005; Vale \& Ostriker 2006; van den Bosch et al. 2007;
Behroozi  et al. 2010; Lu et al 2014), have been developed to understand how galaxies
form and evolve in dark matter halos. However our understanding of the physical
processes driving galaxy formation and evolution is still poor, and many competing
models have been proposed.

It is, therefore, crucial to develop an optimal strategy that can sufficiently utilize the abundant
observational data to constrain theoretical models. One promising way in this
direction is to reconstruct the initial density field from which the observed universe has
evolved, so that we can carry out simulations for the observed structures under various
model assumptions. Such an approach makes it possible to compare model predictions
with data in a way that is free of the cosmic variance produced by large-scale environments.
This is particularly important when the observational samples are small,
so the cosmic variance is a serious issue, as is the case for dwarf galaxies and
for low-$z$ quasar absorption line systems.  Moreover, accurate constrained
simulations can also be used to study the histories and environments of real
galaxies, as well as to investigate the interactions among galaxies, gas and dark matter.

There have been a lot of efforts in developing methods to reconstruct the initial conditions of
the local universe (e.g. Peebles 1989; Hoffman \& Ribak 1991; Nusser \& Dekel 1992; Klypin et al. 2003;
Brenier et al. 2003; Lavaux 2010; Jasche \& Wandelt 2013; Kitaura 2013; Doumler et al. 2013; Wang et al. 2013; 2014; Sorce et al. 2016).
These methods differ primarily in three ways. The first is the observational data that are used to
constrain the reconstruction. Some of the previous investigations use galaxies as tracers of the
cosmological density field, while others use the peculiar velocities of galaxies.
With the advent of large redshift surveys, millions of galaxies now have accurate measurements
for their positions, redshifts and intrinsic properties. Using galaxy distribution to trace the cosmic
density field is an obvious choice.  However, one long-standing problem is that galaxies
are biased tracers of cosmic density, and the bias is not easy to model as it depends on
how galaxies form in the cosmic density field.  In the gravitational instability scenario
of structure formation, the peculiar velocities are directly related to the mass density field.
However, only a small fraction of nearby galaxies have reliable peculiar velocity measurements
(e.g. Willick et al. 1997; Tully et al. 2013). More recently, halos represented by galaxy groups (e.g. Yang et al. 2005; 2007)
have been used to reconstruct the current mass density field (Wang et al. 2009;2013; Mu{\~n}oz-Cuartas et al. 2011), that can also serve as the base of the reconstruction. This method follows the spirit of current halo model (e.g. Cooray \& Sheth 2002), and automatically takes into account the halo bias. The second difference is in the dynamical model
used to predict the final density field from the initial conditions. A number
of dynamical models have been used, including linear theory, Zel'dovich approximation
and its modification (Nusser \& Dekel 1992; Wang et al. 2013; Doumler et al. 2013; Sorce et al. 2016), second-order Lagrangian perturbation theory and its modifications (Jasche \& Wandelt 2013; Kitaura 2013; He{\ss} et al. 2013),  and Particle-Mesh (PM) dynamics (Wang et al. 2014).
The third difference is in the model used to infer the initial conditions from the observational
constraints. Some of the previous investigations adopted backward models which
evolve the present-day density/velocity fields back in time to the linear regime
(Nusser \& Dekel 1992; Doumler et al. 2013; Sorce et al. 2016). Recently, forward models are proposed, which use a probability distribution function consisting of a Gaussian prior and a likelihood function, and
use MCMC-like algorithms to obtain the posterior of the initial density field (Jasche \& Wandelt 2013; Kitaura 2013;  Wang et al. 2013; 2014).
Both the forward and backward models have their shortcomings: the backward models
are only valid on scales where the multi-streaming (shell crossing) is absent,
while the forward models require a large amount of computations to sample
the parameter space. We refer the reader to Wang et al. (2014) for a more detailed
discussion of the various methods.

In our first ELUCID paper (Wang et al. 2014, hereafter Paper I),  we developed a forward method
which employs the Hamiltonian Markov Chain Monte Carlo (HMC) algorithm to sample the posterior
and uses the PM dynamics to evolve the initial condition to the present day.
The reliability and accuracy of the HMC method were examined in great detail in
Paper I and Wang et al. (2013) using simulations and mock catalogs constructed from them. In the second paper (Tweed et al. 2016 in preparation), we use $N$-body simulations to examine the reliability of the HMC method in reproducing the dark matter halo population.
We found that our method can effectively trace massive halos back to the initial condition,
which demonstrates that our reconstructed initial conditions can be used to simulate
the formation histories of individual massive halos.

As the third in a series, this paper presents the application of method to the SDSS data.
The paper is organized as follows. Section \ref{sec_method} describes briefly the method
and observational data we use.  Section \ref{sec_sts} presents the statistical properties of
our constrained simulation. The relations of the mass density field in the constrained
simulation with the SDSS groups and galaxies are investigated in Section \ref{sec_corr}.
Some interesting structures in the cosmic web as revealed by the constrained simulation
are presented Section \ref{sec_web}. Finally, we summarize our results and make
further discussions some in Section \ref{sec_sum}.

\section{Data and Reconstruction Method}
\label{sec_method}

In this section, we present a brief introduction to our reconstruction method developed
in our early studies (Yang et al. 2005;2007; Wang et al. 2009;2012;2013; Paper I).
The method starts from a redshift catalog of galaxies, such as the SDSS.
First, we employ the halo-based group finder to identify galaxy groups from the galaxy catalog. Second, a model based on linear perturbation
theory is applied to predict the peculiar velocity of each halo, and to correct the redshift
to obtain its distance. Third,  the groups in ``real space'' are used to reconstruct
the present-day density field in the local Universe. Finally, we employ a HMC$+$PM
method we have developed to reconstruct the initial conditions
of the density field in the local Universe, and use $N$-body simulations to evolve the
reconstructed initial conditions to obtain the formation history of the local Universe.

\begin{figure*}
\centering
\includegraphics[width=1.0\textwidth]{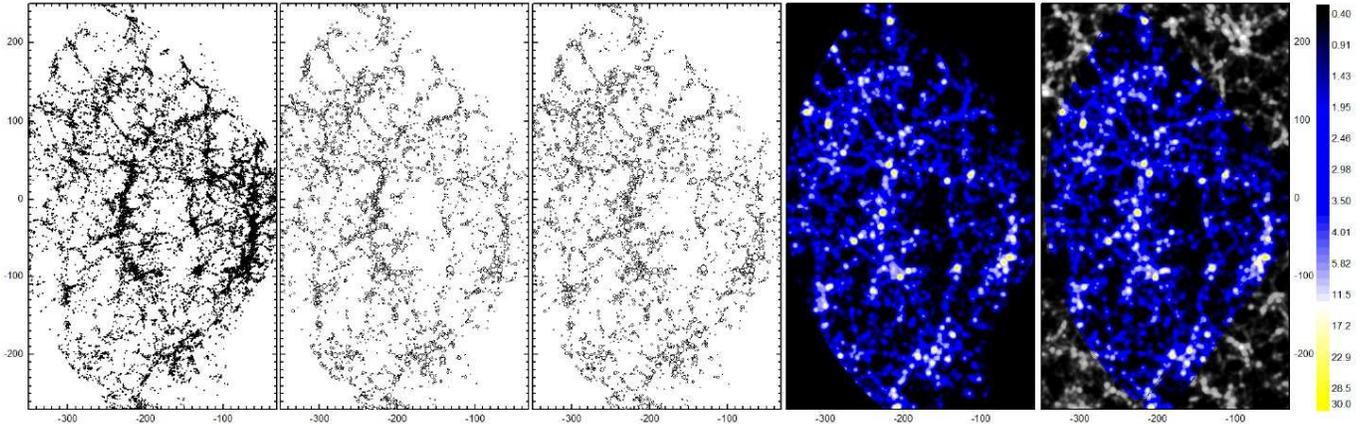}
\caption{Galaxy, Group and Mass distributions in a 10$\mpc$ thick slice containing the Sloan Great Wall.
First panel: galaxy distribution; Second panel: group distribution, the sizes of groups are scaled with their
virial radii; Third panel: group distribution after correcting for redshift distortion; Fourth panel: projected mass
density field of local Universe reconstructed from the group catalog; Fifth panel: projected mass density field
obtained from constrained simulation at $z=0$, and the color-coded region is the reconstruction volume.
The density fields in the two right panels are smoothed with a Gaussian kernel with a radius of
$2.0\mpc$ and scaled with the mean density of the universe. The corresponding values of the densities
are indicated by the color bar.}
\label{fig_drec}
\end{figure*}

\subsection{SDSS galaxies and groups}
\label{sec_gr}

The galaxy sample used here is formed from the New York University
Value-Added Galaxy Catalogue (Blanton et al. 2005)
of the SDSS DR7 (Abazajian et al. 2009), but includes a set of improved reductions
(see Yang et al. 2007). We select all galaxies in the main galaxy sample, with $r$-band
apparent magnitudes $\le17.72$ after correcting
for Galactic extinction, with redshifts in the range $0.01\leqslant z \leqslant 0.2$,
and with redshift completeness $C > 0.7$.
Galaxy groups are identified by using the adaptive halo-based group
finder developed by Yang et al. (2005; 2007). Group masses are estimated
using abundance matching based on the ranking of a characteristic
luminosity/stellar mass defined for each of the groups. Here, we adopt their luminosity-based
group masses. Tests based on detailed mock catalogs suggest that more than 90\% of
the true halos can be identified and the fraction is even higher at the low and high
mass ends. The uncertainties in the group mass estimation is about 0.35 dex for
groups in the intermediate mass range, and 0.2 dex for both low and high mass
groups (see Yang et al. 2007 for details).

The SDSS DR7 covers a sky area of 7,748 deg$^2$, consisting of two
parts: a larger region in the Northern Galactic Cap (NGC)
and a smaller region in the Southern Galactic Cap (SGC). In order to obtain a reliable
reconstruction, we only use the more contiguous NGC region, which has a sky coverage of
7047 deg$^2$.  As shown in Yang et al. (2007), the mass assignment based on
the characteristic luminosity is complete to $z\sim 0.12$ for groups with
mass $M_{\rm gr}\gs 10^{12}\msun$. In the following, we only use groups with
$M_{\rm gr}\geq M_{\rm th}=10^{12}\msun$, so our reconstruction is restricted to
the redshift range $0.01\leq z\leq0.12$, which will be referred to as the survey
volume. Within the survey volume, the total number of groups with masses
above $M_{\rm th}$ is 121,922.

The left two panels of Figure \ref{fig_drec} show the spatial distributions of the
galaxies and groups of $M_{\rm gr}\geq M_{\rm th}$ in a slice of $10\mpc$ in thickness.
The slice is chosen so that a large scale structure, called the Sloan Great Wall
(Gott et al. 2005), is located at the center. One can see that there are lots of stick-like
structures in the high density regions of the galaxy distribution, which point towards to the earth,
which are the `Fingers of God' (FOG) caused by the virial motions of galaxies within galaxy clusters.
Such a FOG effect is largely mitigated in the distribution of groups, since the group
finder takes into account the virial motions when assigning galaxies into groups. The
reader is referred to Yang et al. (2007) for the details about the group catalog construction
and halo mass estimation. Here we adopt the WMAP5 cosmology (Dunkley et al. 2009),
instead WMAP3 adopted in the original Yang et al. (2007) paper. This mean
all quantities are updated with this new cosmology.

\subsection{Correction for redshift distortion}
\label{sec_crd}

Since the method (shown in Section \ref{sec_hd}) used to reconstruct present-day density field from dark matter halos is designed to
work, and has been tested in great detail, only in real space, we need to correct for redshift distortion in the
distribution of groups in order to reconstruct the the present-day density field in real space from groups.
In addition to the virial motion within groups, the large-scale bulk motions of groups
can also produce redshift distortion (Kaiser 1987). To this end,
we use the method developed in Wang et al. (2009; 2012). The basic idea is to apply
the linear theory to the mass density field represented by groups to predict the
peculiar velocity of each group, and then use it to correct for the redshift distortion.
Since the procedure starts from groups in redshift space, iterations are required.
Our tests suggest that two iterations are sufficient to achieve convergence.
Of course, the linear theory is invalid on small scales; in particular it may predict
too high bulk velocities around high density regions, such as clusters of galaxies.
To reduce errors caused by this, the group mass density field is smoothed using a
Gaussian kernel with a relatively large scale, corresponding to a mass
of $\log(M_{\rm s}/\msun)=14.75$ at the mean density. The distribution of groups
in ``real space'' is shown in the third panel of Figure \ref{fig_drec}. Our detailed
tests based on mock group catalog showed that the mean offset between the
predicted and original positions of groups is between 1.1 and 1.4$\mpc/h$ in the
inner region of the survey volume (Wang et al. 2012; 2013). Such spatial offset
corresponds to a velocity of $\sim130\kms$ according to Hubble law, which is similar
to the typical uncertainty in the predicted velocities (e.g. Wang et al. 2009).
In locations close to the boundary, the predicted velocities become more inaccurate,
because velocity field is sensitive to large scale structures that are not well samples
near the boundary.

\subsection{Reconstructing the present-day density field}
\label{sec_hd}

\begin{figure}
\centering
\includegraphics[width=0.5\textwidth]{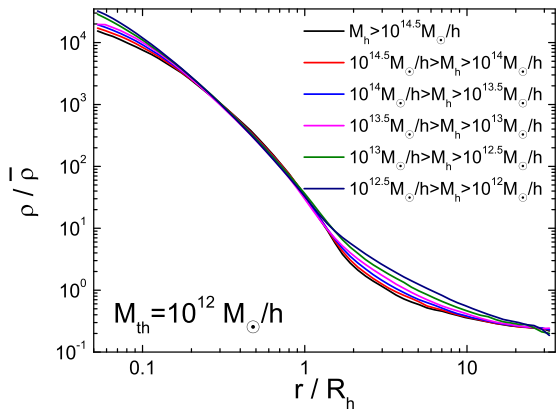}
\caption{Density profile around halos in domain for halos in different mass bins, estimated from simulations}
\label{fig_dp}
\end{figure}

The group catalog in real space described above is used to reconstruct the present-day
mass density field in the local Universe, using a method that closely follows the
halo-domain reconstruction method developed in Wang et al. (2009; 2013).
This method ``convolve'' the dark halo population (as represented by individual
groups) with certain density profiles to reconstruction the large scale density field.
These density profiles are obtained from 8 $N$-body simulations\footnote{Four of them
use $512^3$ particles to trace the evolution of the cosmic density field in a cubic box with
a side length of $300\mpc$, and four use $512^3$ particles in a smaller, $100\mpc$, box.}
of the WMAP5 cosmology using Gadget-2 code (Springel 2005). Following
Wang et al. (2009), we partition each simulation box into a set of domains. Each domain
contains one and only one halo with mass of $M_{\rm h}\geq M_{\rm th}$. The
domain geometry of a halo is determined in such a way that any point in the domain
is closer, based on a specific distance proximity, to the halo than to any other halo.
The distance proximity between a point and a halo is defined as $r/R_{\rm h}$, where
$r$ is the distance between them and $R_{\rm h}$ is the virial radius of the halo.
We then calculate the average density profile around halos of the same mass in
their corresponding domains. Figure \ref{fig_dp} shows the results in six different mass bins.
The eight simulations yield almost identical density profiles; we thus only show the results
averaged over them.

The density field is then reconstructed in the following way. For a SDSS group, we pick a
density profile according to its halo mass. A Monte Carlo technique is used to put particles
around this group to sample the chosen density profile in the three dimensional space
within its domain. This procedure is repeated for all the groups with masses above $M_{\rm th}$.
Eventually, we get a present-day density field in the local Universe, which we denote by
$\rho'_{\rm hd}({\bf x})$. In Wang et al. (2009), the method was applied to dark halos
extracted directly from $N$-body simulations. There it was found that the reconstructed
density field correlates with the original simulated density field
[denoted as $\rho_{\rm si}({\bf x})$] extremely tightly. Wang et al. also applied
the method to mock group catalogs to examine its reliability against uncertainties
arising from false identification of groups and redshift distortion effects
(Wang et al. 2013). It was found that, on scales of $\sim3\mpc$, the root mean
square (r.m.s) of $[\rho_{\rm si}({\bf x})-\rho'_{\rm hd}({\bf x})]/\rho'_{\rm hd}({\bf x})$
is between 0.3 and 0.5, corresponding to a scatter between 0.11 and 0.18 dex.
These tests demonstrate that our method can reliably recover the density field
from groups constructed from a galaxy redshift survey.

However, there is a small but systematical bias in the power spectra of the
reconstructed density fields, as demonstrated in Wang et al. (2009).
When the method was applied to halos extracted from $N$-body simulations,
the power spectra of the reconstructed density fields, $P_{\rm hd}(k)$,
were found to be 10\%-20\% lower than those of the original simulations,
$P_{\rm sim}(k)$, on some scales. This loss of power is due to the fact that we
have ignored the contribution of smaller halos and that the non-spherical
nature of the structures on halo and larger scale is not accurately modeled
by the density profiles, such as those shown in Figure \ref{fig_dp}.
The ratio between these power spectra is about the same for all simulations.
Thus, we can correct for the bias simply by using a correcting function defined as,
$C_{\rm p}(k)=\sqrt{P_{\rm sim}(k)/P_{\rm hd}(k)}$. In practice we proceed
in the following way. We embed $\rho'_{\rm hd}({\bf x})$ in a periodic box of size
$L_{\rm cs}=500\mpc$, which is divided into $N_{\rm c}=500$ grid cells in each
dimension. For the grid cells inside the survey volume, we obtain the over-density,
$\delta'_{\rm hd}({\bf x})=(\rho'_{\rm hd}({\bf x})-\bar{\rho})/\bar{\rho}$,
where $\bar{\rho}$ is the mean density within the survey volume. For grid cells
outside the survey volume, $\delta'_{\rm hd}({\bf x})$ is set to zero.
We then Fourier transform $\delta'_{\rm hd}({\bf x})$ and multiply the
resulting Fourier modes by $C_{\rm p}(k)$. Finally, we obtain the over-density
field ($\delta_{\rm hd}({\bf x})$ ) by the inverse Fourier transformation, and
the densities are obtained through $\rho_{\rm hd}({\bf x})=(\delta_{\rm hd}({\bf x})+1)\bar{\rho}$.
In the forth panel of Figure \ref{fig_drec}, we show the reconstructed density field
smoothed by using a Gaussian kernel of radius $R_{\rm s}=2\mpc$.
One can see that this density field matches well the distributions of both
the galaxies and groups. In the following, we denote this reconstructed
present-day density field by $\rho_{\rm hd}$.

In Appendix, we use realistic galaxy and group mock catalogs to further test
our halo-domain method. We also estimate the uncertainties of the reconstruction
by comparing the reconstruction and the original used to construct the mock catalog.
We find that $\rho_{\rm hd}$ is strongly correlated with the density field of the original
simulation, $\rho_{\rm si}$, without any significant bias. The typical scatter between
the two densities, in terms of the r.m.s of $\log(\rho_{\rm hd}/\rho_{\rm si})$, are
0.09 and 0.20 dex at $R_{\rm s}=4$ and $2\mpc$ respectively. Moreover, we find
that implementing the correction function improves the reconstruction at low density
regions but has no significant effect on high density regions.

\subsection{Reconstructing the initial density field}
\label{sec_ic}

The HMC method (Wang et al. 2013; Paper I) is then employed to reconstruct the initial
density field from the halo-domain density field. This method infers a linear
density field in Fourier space, $\delta({\bf k})$, from a posterior probability
distribution function,
\begin{eqnarray}
\emph{Q}(\delta_j({\bf k})|\rho_{\rm hd}({\bf x}))
&=&{\rm e}^{-\chi^2}\times G (\delta({\bf k}))\nonumber\\
&=&{\rm e}^{-\sum_{\bf x}[\rho_{\rm mod}({\bf x})-\rho_{\rm hd}({\bf
x})]^2\omega({\bf x})/2\sigma_{\rm hd}^2({\bf x})}\times\nonumber\\
&&\prod_{\bf
k}^{\rm half}\prod_{j={\rm re}}^{\rm im}\frac{1}{[\pi P_{\rm lin}(k)]^{1/2}}{\rm
e}^{-[\delta_j({\bf k})]^2/P_{\rm lin}(k)}\label{eq_post}\,.
\end{eqnarray}
The second term, $G (\delta({\bf k}))$, is the prior term that ensures
$\delta({\bf k})$ to have a Gaussian distribution with a given linear power
spectrum, $P_{\rm lin}(k)$. The likelihood term, ${\rm e}^{-\chi^2}$, is designed
to ensure that the predicted density field ($\rho_{\rm mod}$) evolved from
$\delta({\bf k})$ according to a given structure formation model matches the
input nonlinear density field, which in our case is the halo-domain density field
$\rho_{\rm hd}$. In our application, $\sigma_{\rm hd}({\bf x})$
is set to $0.5\rho_{\rm hd}({\bf x})$, and the weighting function,  $\omega({\bf x})$, is specified
below.

Since the input final density field is highly non-linear (Figure \ref{fig_drec}),
it requires the structure formation model that links $\delta({\bf k})$ and $\rho_{\rm mod}({\bf x})$
to be sufficiently accurate so as to properly handle highly nonlinear dynamics on small scale.
As shown in Paper I, an inaccurate model can introduce spurious
non-Gaussian features in the reconstructed linear density field, which in turn
can lead to significant bias in the predicted halo population. In this paper, we
will follow Paper I and use the Particle-Mesh (PM) dynamics to follow the
evolution of the density field. The PM scheme is often used to calculate the
forces on large-scales  in cosmological simulations.  It can describe the structure
formation and evolution accurately as long as the available computational resources
allow the implementation with sufficiently high resolutions in both space and time.
In what follows, reconstruction method based on the HMC with PM dynamics
is referred to as the HMC$+$PM method.

Some important parameters used in the HMC and PM have to be specified before
using the HMC$+$PM method to reconstruct the initial density field.
As discussed in the previous section, the halo-domain density field  is embedded
in a 500$\mpc$ cubic box and the box is divided into $500^3$ grid cells, so that the
grid cell size is $l_{\rm c}=1\mpc$. For such a spatial resolution, twenty PM time steps
are sufficient (see Paper I for detailed discussions). Since we perform the
reconstruction in a very large volume, a more accurate PM model (equivalent to
higher spatial resolution) will require computational resources that are beyond our
reach at the moment. We note, however, that the PM model with
$l_{\rm c}=1\mpc$ and 20 time steps is already much more accurate than any other
structure formation models used so far in the literature for initial density field
reconstruction, such as the Zel'dovich approximation and higher-order Lagrangian
perturbation theory. With such a PM, significant discrepancies are still present
on small scales between the model prediction and the fully non-linear
density field, due to the inaccuracy of the model. The tests in Paper I
suggest that smoothing both the density fields on a scale of $\geq 3l_{\rm c}$
can effectively eliminate such discrepancies. Here, we choose a more conservative
smoothing scale of $4l_{\rm c}=4\mpc$.

Another important parameter is the weighting function $\omega({\bf x})$, which
specifies the window in which the comparisons between $\rho_{\rm mod}$
and $\rho_{\rm hd}$ are to be made. In our reconstruction, all grids outside of
the survey volume are assigned a zero weight, but not all grids inside the survey
volume are assigned a full weight. Recall that the halo-domain density field,
$\rho_{\rm hd}$, is smoothed on $R_{\rm s}=4\mpc$ in the HMC run. When
performing the smoothing, the density outside of the survey volume is set to the
mean cosmic density. It thus introduces uncertainties into $\rho_{\rm hd}$ at
locations close to the boundary of survey volume. To deal with this boundary
effect, we define a factor $f_{\rm c}(R_{\rm b})$ to characterize the closeness
of a grid to the boundary. For a grid cell in the survey volume, $f_{\rm c}$ is
the fraction of a spherical volume of radius $R_{\rm b}$ centered on the grid that
is contained by the survey volume. Here we choose $R_{\rm b}=10\mpc$,
and use $f_{\rm c}(10\mpc)=0.84$ to define a boundary layer.
When the curvature radius of the boundary surface is much larger than $10\mpc$
(which is the case for most of the cases), $f_{\rm c}=0.84$ corresponds to
distance of $\sim5\mpc$ from the boundary,  which is larger than the adopted
smoothing scale. So the boundary effect can be ignored for all the grids
with $f_{\rm c}(10\mpc)>0.84$.  Such grids are assigned a weight of unity, while
the grids with $f_{\rm c}(10\mpc)\leq0.84$ are assigned a weight of zero.
The final volume (reconstruction volume hereafter) is about $2.73\times10^{7}(\mpc)^3$,
about 12\% smaller than the original survey volume.
Other parameters used in the HMC are set to be the same as in Paper I,
and will not be repeated here.

\begin{figure}
\centering
\includegraphics[width=0.5\textwidth]{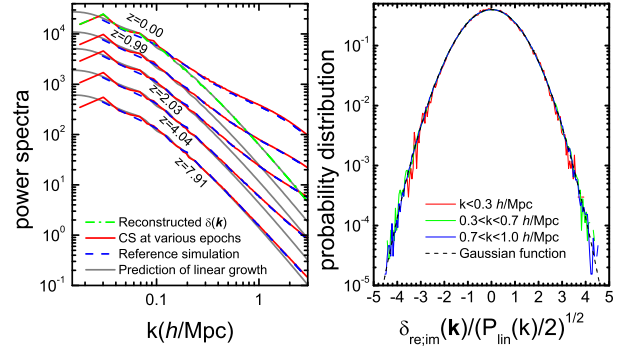}
\caption{Left panel: the power spectra of the reconstructed linear density field (green dash-dotted line) and from
the constrained simulation at various epochs (red solid lines). We also show the analytical linear power spectrum
(grey lines) and those of the reference simulation (blue dashed lines) for comparison.
Right panel: the probability distributions of the real and imaginary parts of the reconstructed linear density field,
normalized by the linear power spectrum, at three different wavenumbers as indicated in the panel.
The dashed line shows the expected Gaussian distribution.}
\label{fig_ps}
\end{figure}

Our HMC$+$PM makes calculations of a total 3000 HMC steps. The HMC system
evolves quickly at the beginning, and achieves a convergent state after about 600 HMC steps.
The typical scatter between $\rho_{\rm mod}$ and $\rho_{\rm hd}$ [hereafter TS, defined as
the r.m.s of $\log(\rho_{\rm mod}/\rho_{\rm hd})$] after the 1000th step remains
around 0.015 dex, indicating that after convergence the model density field match
the input constraint very well.  After convergence, the statistical properties of
$\delta({\bf k})$ remain similar, so we only show the results at the 1500th step.
The green dash-dotted line in the left panel of Figure \ref{fig_ps}
shows the power spectrum of $\delta({\bf k})$ at the 1500th HMC step, which is in good
agreement with the theoretical power spectrum used in the model.
The right panel shows the probability distributions of the real and imaginary parts of $\delta({\bf k})$
for three different wave-numbers. These distributions closely match a Gaussian distribution,
demonstrating that our method recovers the Gaussian property of the linear density
field very well.

\subsection{The constrained simulation}
\label{sec_resimu}

The linear density field output at the 1500th step is chosen to set up the initial condition for
an $N$-body simulation. The initial condition is sampled with $3072^3$ dark matter particles
at an initial redshift of $z_{\rm ini}=100$. The mass of each particle is $3.09\times10^8\msun$.
Because the HMC process only generates the Fourier models at
$k\leq N_{\rm c}\pi/L_{\rm cs}$, the missing modes at larger $k$ are complemented by
assigning random phases. The density field is then evolved to present day
using L-GADGET, a memory-optimized version of GADGET-2 (Springel et al. 2005).
One hundred snapshots from $z=19$ to $z=0$ are produced, with the expansion
factor evenly spaced in logarithmic space between successive snapshots.
The simulation obtained this way is referred to as the constrained simulation
(hereafter CS) in the following.

In the fifth panel of Figure \ref{fig_drec}, we show the present-day density field of the
constrained simulation, $\rho_{\rm cs}$. Within the reconstruction volume, almost all
large-scale structures, and even some small structures, are well recovered. The typical
scatter (TS) between $\rho_{\rm hd}$ and $\rho_{\rm cs}$ is 0.018 dex when smoothed on
a scale of $R_{\rm s}=4\mpc$, only slightly larger than the TS between $\rho_{\rm hd}$
and $\rho_{\rm mod}$. When the densities are smoothed on $2\mpc$, the TS increases
to about 0.13 dex. Given the large dynamical range in density involved,
about 2 orders of magnitude, this relatively low TS indicates that
$\rho_{\rm cs}$ matches $\rho_{\rm hd}$ well even on scales down to
$\sim 2\mpc$. Note, however, the scatter discussed here only represents the
uncertainties arising from the HMC$+$PM method; it does not include
uncertainties caused by the other parts of the reconstruction discussed
above. To estimate the full uncertainties of the constrained simulation, we have applied
the reconstruction method to a realistic galaxy mock catalog (see Appendix).
The TS between $\rho_{\rm cs}$ and the true density $\rho_{\rm si}$ are 0.10 and 0.23 dex
for $R_{\rm s}=4\mpc$ and $2\mpc$, respectively.
Recall that the TS between  $\rho_{\rm hd}$ and $\rho_{\rm si}$ are 0.09 dex and 0.20 dex
for the two values of $R_{\rm s}$, respectively. This suggests that the major sources
of the uncertainties are from the inaccuracies of the redshift-distortion correction,
group finder, and halo-domain reconstruction. See the Appendix for detailed discussions.

\section{General statistical properties of constrained simulation}
\label{sec_sts}

In this section, we present some statistical properties of the constrained simulation.
When necessary, the results obtained from an $N$-body simulation with random initial
phases are used for comparisons. The reference simulation assumes the same cosmology
as our constrained simulation and its density field is traced by $1024^3$ particles in a
cubic box of 300$\mpc$ on a side.

Power spectra are often used to measure the mass clustering and to characterize the
large scale structure.  In Figure \ref{fig_ps}  the power spectrum of the reconstructed linear
density field is shown together with the power spectra of the constrained simulation (CS)
at five different redshifts.  The power spectra predicted by the linear pertubation theory
and those of the reference simulation are also shown for comparison. As one can see,
at the linear scales ($k<0.15\mpchi$), the CS power spectra match the linear spectra very well.
At smaller scales, the CS power spectra are significantly higher than the linear spectra,
and the difference increases with the decreasing redshift, clearly owing to non-linear
evolutions of the density field.  However, these CS spectra match with
those of the reference simulation almost perfectly,  suggesting that the constrained
simulation reproduces well the large scale structures statistically.

\begin{figure}
\centering
\includegraphics[width=0.5\textwidth]{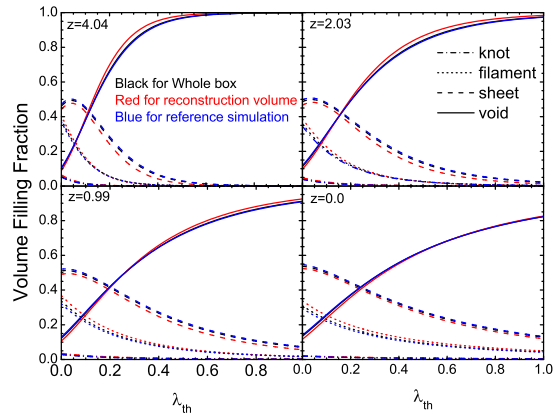}
\caption{The fractions of volume classified as knots (dot-dashed lines), filaments (dotted lines),
sheets (dashed lines) and voids (solid lines) as a function of the threshold $\lambda_{\rm th}$ at
four different redshifts. }
\label{fig_morth}
\end{figure}

Since morphology of the cosmic web is complicated (Figure \ref{fig_drec}),
the power spectrum alone is not sufficient to completely characterize a cosmic density
field. Recently, much effort has been giving to developing methods to classify the
cosmic web (e.g. Hahn et al. 2007; Lavaux \& Wandelt 2010; Hoffman et al . 2012; Leclercq et al. 2016). Here we employ a dynamical classification, first developed
by Hahn et al. (2007) and refined by Forero-Romero et al. (2009). The basic
idea is to use the eigenvalues of the tidal tensor (or deformation tensor) to
define the type of the local structure in a cosmic web. The tidal tensor, ${\cal T}_{ij}$,
is defined as
\begin{equation}\label{eq_tij}
{\cal T}_{ij}=\partial_i\partial_j\phi\,,
\end{equation}
where, $\phi$ is the peculiar gravitational potential and can be
calculated from the CS density perturbation ($\delta_{\rm cs}$) through
a modified Poisson equation:
\begin{equation}
\nabla^2\phi=\delta_{\rm cs}
\label{eq_phi}\,.
\end{equation}
Note that this equation is scaled by $4\pi G\bar{\rho}$ and $\delta_{\rm cs}$ is
defined over gird cells and smoothed on a comoving scale of $2\mpc$.
We derive the three eigenvalues of the tidal tensor, $\lambda_1\geq\lambda_2\geq\lambda_3$,
at each grid point, and classify a grid point by counting the number of eigenvalues
above a given threshold $\lambda_{\rm th}$. A knot point corresponds to
all the three eigenvalues being $>\lambda_{\rm th}$, while a filament point to two,
a sheet point to one, and a void point to zero.

The classification of the cosmic web at a given point depends on the choice of
$\lambda_{\rm th}$. It is still unclear which threshold value is the best. In Figure \ref{fig_morth},
we show the ratios of the volumes in the four types of structures to the whole volume
(the volume filling fraction) as functions of $\lambda_{\rm th}$ at four different redshifts.
Generally, the volume filling fractions for knots, filaments and sheets decrease with
increasing $\lambda_{\rm th}$, while that for voids exhibits an opposite behavior.
As discussed in Forero-Romero et al. (2009), the eigenvalues
of the tidal tensor are not only dynamical quantities, but also reflect
the collapse timescale along the corresponding eigenvectors.
Thus, these results reflect the collapse time distributions of the structures in three
dimensions.  Interestingly, the volume filling fraction defined with  $\lambda_{\rm th}=0$
show almost no variation with redshift. Theoretically,  $\lambda_{\rm th}\sim0$ corresponds
to a very long collapse time, longer even than the Hubble time (Forero-Romero et al. 2009).
It is thus not surprising that the fractions at $\lambda_{\rm th}=0$ are
independent of redshift. However, the dependence of the filling fraction
on $\lambda_{\rm th}$ varies significantly with redshift: the
change tends to occur in a narrower range of $\lambda_{\rm th}$
at higher redshift.  By definition $\lambda_1+\lambda_2+\lambda_3\equiv\delta_{\rm cs}$.
Since the amplitude of the density fluctuation decreases with increasing redshift,
the range of the three eigenvalues also decreases with increasing redshift, and
so the behavior with $\lambda_{\rm th}$ is expected.
The results for the reference simulation are presented for comparison.
The fraction of filament (sheet) points in the constrained volume is slightly larger
(smaller) than that in the whole box as well as in the reference simulation,
likely because of cosmic variance. Overall, the average properties of the
cosmic web in the CS are very similar to those of the reference simulation.
More discussions about the cosmic web and its classification will be
presented later (Section \ref{sec_web}).

\begin{figure}
\centering
\includegraphics[width=0.5\textwidth]{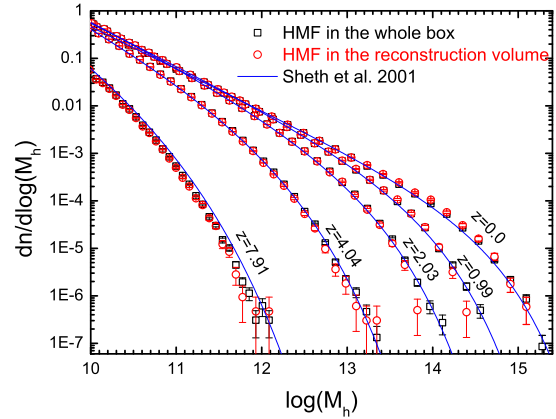}
\caption{The halo mass functions at five different redshifts as indicated in the figures.}
\label{fig_hmf}
\end{figure}

Embedded in these large structures are virialized dark matter halos formed through non-linear
gravitational collapse. In the current paradigm of galaxy formation, halos are thought to be
the hosts within which galaxies form and reside. They thus provide a key link between
galaxies and the cosmic density field (e.g. Mo et al. 2010). Understanding the properties
of the halo population, therefore, is an important step towards understanding
galaxy formation. Furthermore, as demonstrated in  Paper I,
if the structure formation model employed in the HMC is not sufficiently accurate, it
can induce non-Gaussian properties in the initial condition,
which in turn can cause the predicted halo mass function to differ from
what one expects to get from the model.  Thus, investigating the properties of the halo
population provides additional test of the reconstruction method. We select halos
from each snapshot with the standard friend-of-friends (FoF) algorithm
(Davis et al. 1985) with a link length equal to 0.2 times the average particle separation.
The halo mass,  $M_{\rm h}$, is the sum of the masses of all particles in the halo.
Figure \ref{fig_hmf} shows the halo mass functions in both the whole box and
the reconstruction volume at five different epochs. Our results excellently match the
theoretical halo mass functions given by Sheth et al. (2001), except for the highest
redshift where the theoretical model  predicts higher number density of halos.
This discrepancy at high-$z$ is very similar to the finding of Reed et al. (2003) in their
(un-constrained) simulations, suggesting that it is due to the inaccuracy of the
model at high $z$.

\begin{figure}
\centering
\includegraphics[width=0.5\textwidth]{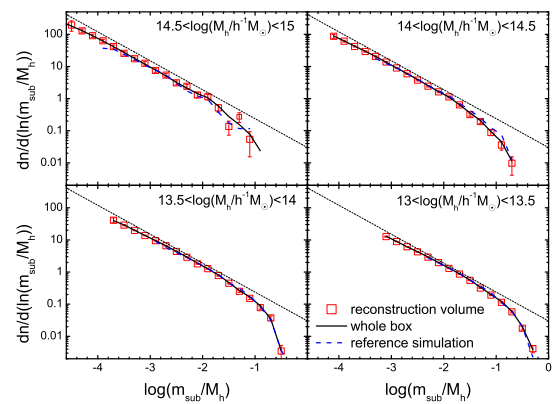}
\caption{Sub-halo mass functions at redshift zero for four host halo mass ranges. The dotted lines show a slope of $-0.90$.}
\label{fig_submf}
\end{figure}

\begin{figure}
\centering
\includegraphics[width=0.5\textwidth]{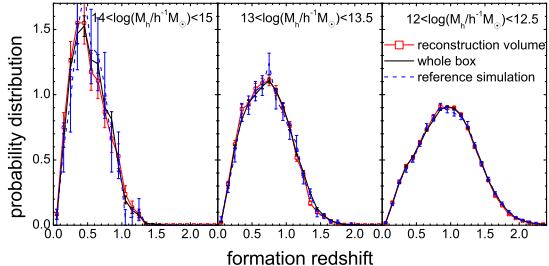}
\caption{The probability distributions of the formation redshift of halos in three different mass ranges.}
\label{fig_fz}
\end{figure}

Halos grow hierarchically, with small halos forming first and merging to form large
halos. After merger, the small halos may become substructures of the large one,
producing a population usually called sub-halos. Host halos/subhalos and their merger
histories extracted from $N$-body simulations have served as the backbone for
semi-analytical model of galaxy formation (Kang et al. 2005; Croton et al. 2006;
Guo et al. 2011; Lu et al. 2011) and other empirical models, such as subhalo abundance
matching (Kravtsov et al. 2004; Vale \& Ostriker 2006; Behroozi  et al. 2010).
To test the ability of our constrained simulation in reproducing the sub-halo population,
we have identified sub-halos in all the FoF halos and constructed merger trees for
all halos and subhalos by using SUBFIND and other relevant codes kindly provided by
Springel (Springel et al. 2005). In Figure \ref{fig_submf}, we show the sub-halo
mass function for host halos in four mass ranges. Only sub-halos within the virial
radii of their host halos and with more than 50 particles are counted. The mass
functions have slopes very close to $-0.9$, in good agreement
with results obtained previously (e.g. Gao et al. 2004; Diemand et al. 2004).
The results for the reference simulation are also presented, both the amplitudes
and shapes are in good agreement with the results from the CS. The
full structure of a merger tree is complex and not easy to be presented in a
simple way. Here we track the main branch (i.e. the most massive progenitor)
of a $z=0$ halo back in time until the time when the mass of the most massive progenitor
is half of the final halo mass. The corresponding redshift, sometimes referred to
as the formation redshift, is often used to represent the formation history
of the halo. We show in Figure \ref{fig_fz} the distributions of the formation
redshift for halo in three mass ranges. One can see that the distributions
obtained from the  CS match well those for reference simulation.

As a summary of this section, the statistical properties of our CS are very similar
to those of general, unconstrained simulations, demonstrating that no
significant bias is introduced by our reconstruction.
With a resolution about 3 times as high as the widely-used Millennium
Simulation (Springel et al. 2005),  this simulation itself can be used to study many
properties of the cosmic web (density, velocity and tidal fields, etc.), to construct
mock catalogs of galaxies, to model galaxy-galaxy lensing, to study halo formation
and distribution,  and to generate halo merger trees to model galaxy formation.
More importantly,  our constrained simulation is unique in that it can be used
to study directly the formation of the structures we observe, and the relation
between the mass density field and real galaxies. Furthermore, it can also be
used to set up zoom-in simulations in which the zoom-in region to be simulated
with high resolutions needs to be designed so as to avoid contaminations by
outside low-resolution particles during the evolution of the density field.

\section{Correlations of constrained simulation with real groups and galaxies}
\label{sec_corr}

\subsection{Correlation between real groups and halos in the constrained simulation}

\begin{figure*}
\centering
\includegraphics[width=1.0\textwidth]{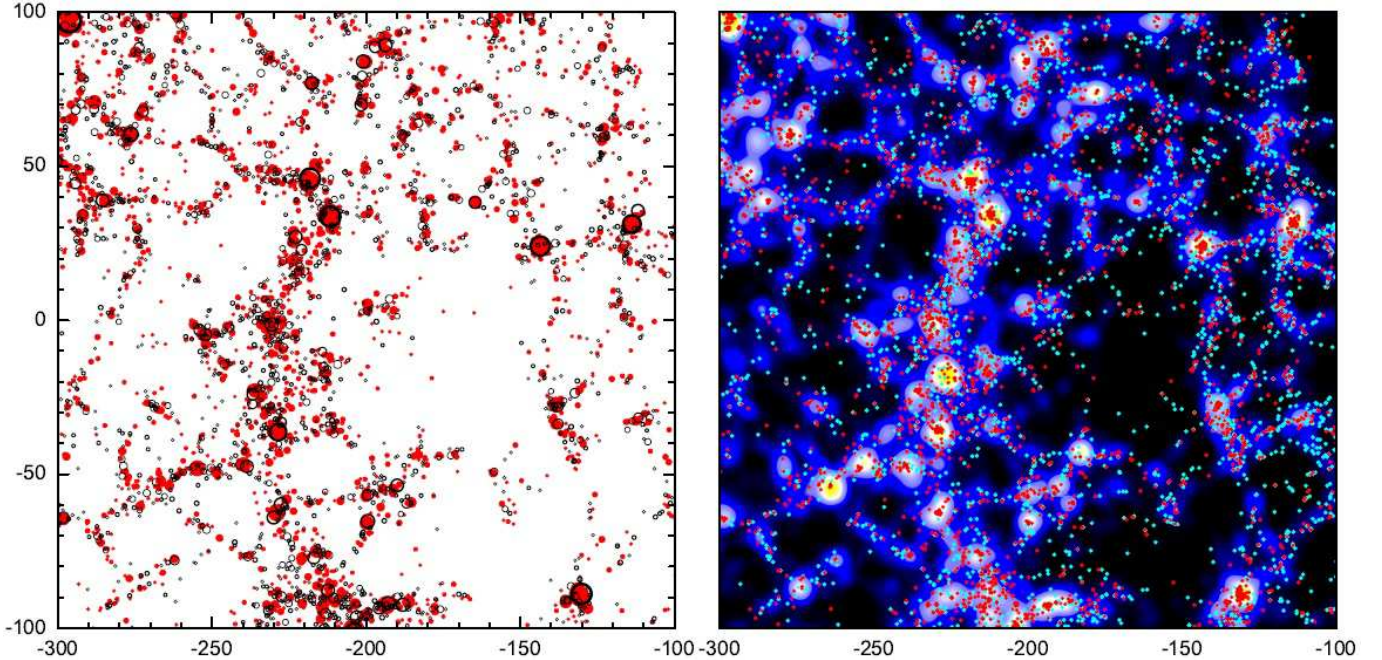}
\caption{Left panel: the spatial distributions of SDSS galaxy groups (red solid circles) and dark halos in the CS
(black open circles) in a 10$\mpc$ thick slice. Only groups and halos with masses larger than $10^{12}\msun$
are shown. The symbol sizes of the groups and halos are proportional to their virial radii. Right panel: the mass density
field together with the galaxy distribution in the same slice as in the left panel. The red (cyan) dots represent
the red (blue) galaxies. The color code for the density field is the same as that in Figure \ref{fig_drec}.}
\label{fig_dggh}
\end{figure*}

Since our reconstruction is based on the group catalog, it is expected that the CS
density field to be tightly correlated with the spatial distribution of galaxy groups
(Figure \ref{fig_drec}). Here we attempt to look for one-to-one correspondences
between the groups and the FoF halos identified from the CS. Before doing this,
let us first look at the spatial distributions of groups and halos. The left panel of
Figure \ref{fig_dggh} shows the distributions of groups and halos with masses above
$M_{\rm th}=10^{12}\msun$ in a $200\times200\times10\mpc$ slice enclosing
the Sloan great wall. On large scales, the spatial distribution of CS halos follows
that of the groups very well. Inspecting individual groups, we can find that almost
all of the massive groups (usually galaxy clusters) have corresponding FoF halos,
However, their sizes and locations do not match perfectly.  For systems of lower
masses, the match becomes worse, but they still appear to reside in similar
large scale structures.

To match groups and halos in a more quantitative way, we construct a galaxy group list and
a halo list. The group list contains all groups with $M_{\rm gr}\geq M_{\rm th}$ in the
reconstruction volume in decreasing order of group mass. Note that the redshift distortion
for these groups has been corrected (Section \ref{sec_hd}). The halo list contains all
FoF halos in the CS (with no limit on halo mass). For the first group (the most massive one)
in the group list, we search in the halo list for halos with distances to the group less
than $\Delta{\rm d}_{\rm mx}$ and with mass satisfying
$|\log{M_{\rm gr}/M_{\rm h}}|\leq \Delta{\rm m}_{\rm mx}$. Here $M_{\rm gr}$ and $M_{\rm h}$
are group and halo masses, respectively. If no halo is found to satisfy the two criteria,
the group is considered to have no match. Otherwise, we choose the halo with the
smallest $|\log{M_{\rm gr}/M_{\rm h}}|$ as the corresponding halo of the group.
Finally the group is removed from the group list,  and the corresponding halo
from the halo list. The procedure is repeated for all groups. This procedure assumes
that large groups have more priority in match a halo than small groups,
and mass matching has more priority than distance matching.
The groups that have corresponding halos are referred to as matched
groups, otherwise unmatched groups.

\begin{figure}
\centering
\includegraphics[width=.5\textwidth]{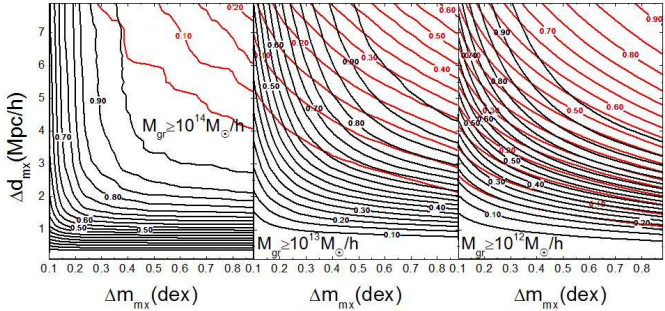}
\caption{The black lines exhibit the cross-matching between galaxy groups and CS halos as a function of the maximum
distance ($d_{\rm max}$) and maximum mass difference ($m_{\rm dex}$) for three group mass ranges.
See the texts for details. The red lines show the results for uncorrelated halo sample, which is generated by rotating the
CS halos with a large angle.}
\label{fig_gf}
\end{figure}

The number of matched pairs depends on the two free parameters,
$\Delta{\rm d}_{\rm mx}$ and $\Delta{\rm m}_{\rm mx}$. In Figure \ref{fig_gf}, we show
the fraction of matched groups as a function of both $\Delta{\rm d}_{\rm mx}$
and $\Delta{\rm m}_{\rm mx}$. Not surprisingly, as $\Delta{\rm d}_{\rm mx}$ and
$\Delta{\rm m}_{\rm mx}$ increase, the number of matched groups increases.
About half of the groups with $M_{\rm gr}\geq10^{14}\msun$ have
matched halos with distances less than $1\mpc$ and $|\log{M_{\rm gr}/M_{\rm h}}| <0.15$.
For $\Delta{\rm d}_{\rm mx}=4\mpc$ (the smoothing scale adopted in our HMC$+$PM)
and $\Delta{\rm m}_{\rm mx}=0.5$, then more than 95\% of massive groups have halo
matches. For comparison, we generate an uncorrelated halo sample by rotating the
CS halos with a large angle. The matching results for the rotated halo sample are
shown in red contours in Figure \ref{fig_gf}. One can see that the fraction of
groups with matched halos is much smaller than 5\% for the same two parameters.

For groups with masses $M_{\rm gr}\geq10^{13}\msun$, the matching
becomes worse. For $\Delta{\rm d}_{\rm mx}=4\mpc$, half of the groups
have halo matches with $\Delta{\rm m}_{\rm mx}=0.25$, and about
80\% of the groups have halo matches with $\Delta{\rm m}_{\rm mx}=0.5$.
The corresponding fractions for the rotated halo sample are only 5\% and 10\%,
respectively.  For groups with $M_{\rm gr}\geq10^{12}\msun$, about 85\% of
groups are matched with CS halos with $\Delta{\rm d}_{\rm mx}=4\mpc$
and $\Delta{\rm m}_{\rm mx}=0.5$. This does not mean that the matching
is better for small halos, however,  since for such a mass
about 40\% of the groups have matched halos in the rotated sample. Small halos are
also much more abundant, so the probability for a group to be matched
with an uncorrelated halo is larger.

\begin{figure}
\centering
\includegraphics[width=0.5\textwidth]{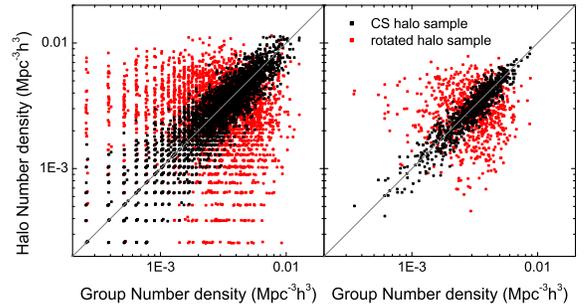}
\caption{The number densities of CS halos (black dots) and rotated halos (red dots) versus the group
number density estimated in cubes of $L_{\rm cb}=20\mpc$ (left panel) and 30$\mpc$(right panel).
Only halos and groups with masses between $10^{12}\msun$ and $10^{13}\msun$ are taken into account.
The solid gray lines indicate the perfect relation.}
\label{fig_ngh}
\end{figure}

For low mass groups, the match with CS halos is only slightly better
than with uncorrelated halos. This suggests that the small scale modes in the
initial condition are only partly recovered. However, as one can see from Figure \ref{fig_dggh},
small groups and CS halos also trace the same large-scale structures.
This indicates that the statistical properties of the small groups are
strongly correlated with small halos in the CS. In order to quantify such correlations,
we compare the correlation between the number density of small groups
and that of CS halos on different scales. To this end, we divide the CS box into
cubes each with a size of $L_{\rm cb}$ on a side, and compute the number densities
of groups, CS halos and halos in the rotated sample in each of the cubes.
Figure \ref{fig_ngh} shows the density - density relation between groups
and halos with masses in the range $10^{12}$ - $10^{13}\msun$ for
cubes with $L_{\rm cb}=20\mpc$ (left panel) and $30\mpc$ (right panel), respectively.
Here only cubes with at least 90\% of their volumes are contained in the
reconstruction volume are included.  As one can see,  the number densities of
CS halos are tightly correlated with those of groups without any significant bias,
and the correlation becomes tighter as $L_{\rm cb}$ increases. In contrast,
no visible correlation can be seen between groups and halos in the rotated sample.

\subsection{Correlation between real galaxies and mass in the constrained simulation}

It is well known that galaxies are biased tracers of the cosmic density field
(e.g. Peacock 1997; Jing et al. 1998). The bias is found to depend on both galaxy
luminosity and color (Norberg et al. 2001; Zehavi et al. 2002;
Li et al. 2006). Thus, the study of the correlation between galaxies
and mass contains important information about galaxy formation.
Almost all of the previous studies are concentrated on statistical measurements,
such as (cross) correlation functions of galaxies. Here we present a one-to-one
comparison between galaxy and mass densities sampled on a grid of cube.

To facilitate comparisons with the CS, we need to correct for the redshift distortions
for galaxies. Our correction method consists of two steps. We first assign the group redshifts (after correction, see Section \ref{sec_crd}) to their member galaxies. Then for each satellite galaxy, we assign a random offset relative to its group center, along the line of sight, based on NFW profile (Navarro et al. 1997) with mass-concentration relation given by Zhao et al. (2009). The method can correct for both Kaiser effect and FOG effect, and make the spatial distribution of satellites within halos follow the NFW profile (see Shi et al. 2016 for more details). The right panel of Figure \ref{fig_dggh} shows the CS density field and galaxy
spatial distribution. We separate red and blue galaxies  according to their binormal
distribution in the $g-r$ color - absolute magnitude plane (e.g. Baldry et al. 2004),
using the same separation line as proposed by Yang et al. (2009):
\begin{equation}
g-r = 1.022-0.0651x-0.00311x^2, x=M_r-5\log{h}+23\,,
\end{equation}
where $M_r-5\log{h}$ are the absolute magnitude in the $r$ band. Note that both
$M_{\rm r}-5\log{h}$ and $g-r$ are $K-$corrected and evolution-corrected to
redshift $0.1$. The galaxy distribution match the CS density field very well.
In addition, red galaxies appear more tightly associated with high density
regions than blue ones.

\begin{figure}
\centering
\includegraphics[width=.5\textwidth]{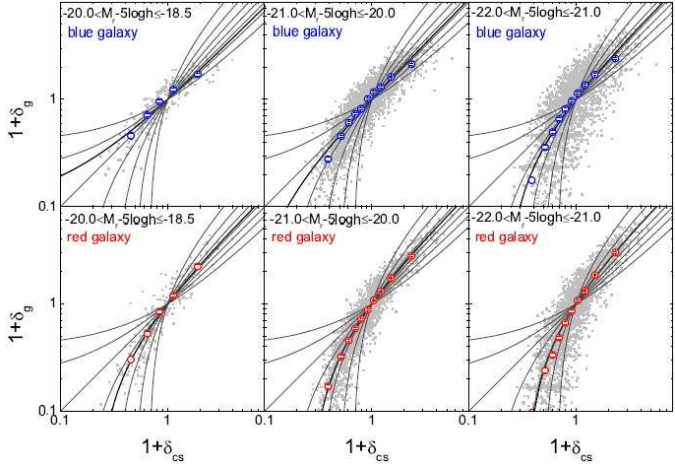}
\caption{The correlations between galaxy densities and mass densities of the CS for galaxies of
different magnitudes and colors,  as indicated in the panels. $\delta_{\rm g}$ and $\delta_{\rm cs}$ are
density contrasts for galaxies and mass, respectively. Each dot point represents the mean
density in a cubic box with $L_{\rm cb}=20\mpc$. The open circles with error bars are the mean values of
the dot points. The thin lines show  $\delta_{\rm g}=b\delta_{\rm cs}$, with $b=$ 0.6, 0.8, 1.0, 1.2, 1.5, 2.0,
and 3.0 (from left to right at $1+\delta_{\rm cs}<1$), respectively.
The thick lines are the best fits to the mean correlations.}
\label{fig_gd20}
\end{figure}

\begin{figure}
\centering
\includegraphics[width=.5\textwidth]{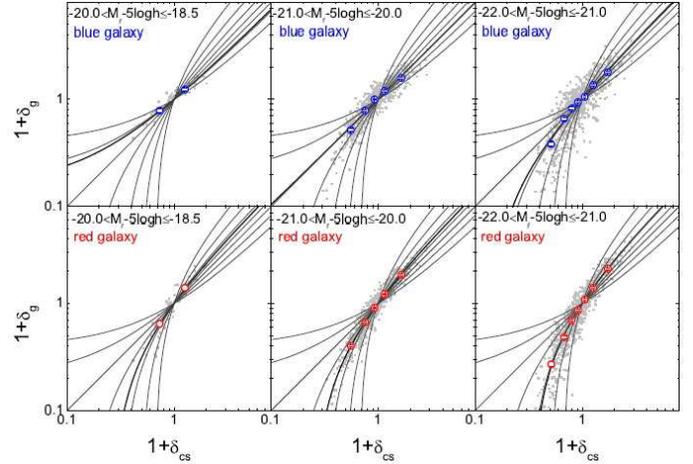}
\caption{Similar to Figure \ref{fig_gd20} but here the densities are the averages in cubic boxes of $30\mpc$ on a side.}
\label{fig_gd30}
\end{figure}

To quantify the correlations in more detail, we select galaxies in three
$M_r-5\log{h}$ bins, $[-18.5,-20)$, $[-20,-21)$ and $[-21,-22)$, and
separate each of them into red and blue subsamples.  We estimate the galaxy density
field for each galaxy subsample as follows. The galaxy subsample is embedded in
the CS box and converted into a density field on a grid by using the clouds-in-cells
(CIC) assignment (Hockney \& Eastwood 1981).
For a given subsample, we estimate the limiting redshift of  completeness
($z_{\rm c}$) using equation (6) in Yang et al. (2009) and the corresponding
co-moving distance $r(z_{\rm c})$. The intersection between the reconstruction
volume and the sphere with radius $r(z_{\rm c})$ is referred to as the complete
volume.  Here, again, we divide the complete volume into equal-sized cubes.
For each cube, the galaxy density contrast is defined to be
$\delta_{\rm g}=\rho_{\rm g}/\bar\rho_{\rm g}-1$, where $\rho_{\rm g}$ is the
galaxy density in the cube and $\bar\rho_{\rm g}$ is the mean galaxy density in
the complete volume.  Using the same method, we also get the mass density
contrast in each cube, $\delta_{\rm cs}$. In the following, we only consider the
cubes that each have more than 90\% of their volumes contained in the
complete volume.

Figures \ref{fig_gd20} and \ref{fig_gd30} show $1+\delta_{\rm g}$
versus $1+\delta_{\rm cs}$ on two different scales, $L_{\rm cb}=20$
and $30\mpc$, for the six galaxy subsamples defined above.
Since the complete volume is different for different sub-samples, the
number of cubes used varies. The galaxy density is strongly correlated with
the CS mass density, and the scatter decreases with increasing $L_{\rm cb}$.
Moreover, the galaxy density is biased with respect to the underlying mass
density. To quantify this, we split each sample of cubes
into several equal-sized subsamples according to $\delta_{\rm cs}$, and
compute the mean $1+\delta_{\rm cs}$ and $1+\delta_{\rm g}$ for each
subsample. The mean correlations are shown as open circles in the
figures. The galaxy bias is usually interpreted in terms of halo bias
(e.g. Mo et al. 1996). We thus use the bias relation $\delta_{\rm g}=b\delta_{\rm cs}$ to fit the
mean results, and show the best-fitting models in thick lines in the figure. The corresponding bias
parameters are list in Table~\ref{tab_bias}. The relations with $b=0.6$, $0.8$, $1.0$,
$1.2$, $1.5$, $2.0$, and $3.0$ are also plotted as references.

In general, the mean correlations can be well fitted by the theoretical model, and the bias factor
tends to be larger for higher luminosity and for red galaxies. In particular, the bias factors obtained
here  match well those obtained by Zehavi et al. (2011), who measured the bias as the square root of
the ratio of the two-point correlation functions between galaxies and dark matter assuming a
fiducial cosmological model.  Inspecting closely the fitting curves, we find that the single-bias model
fit the observational data better for red galaxies than for blue galaxies. Moreover, for a given luminosity,
the correlation also appears tighter for red galaxies. One possible explanation for these results is
that, for a given luminosity,  blue galaxies tend to reside in halos with a wider range of halo masses
than red galaxies.  In a future paper, we will investigate the correlation between the galaxy population and
the reconstructed density and tidal fields in detail.

\begin{table}
\begin{center}
\caption{Galaxy bias for various galaxy subsamples}
\label{tab_bias}
\begin{tabular}{lccc}
\hline
&& $L_{\rm cb}=20\mpc$& \\
\hline
$M_{\rm r}$ bin & $[-18.5,-20)$ & $[-20,-21)$ & $[-21,-22)$ \\
 \hline
blue  & $0.90\pm0.03$ & $1.08\pm0.01$ & $1.29\pm0.01$\\
red  & $1.29\pm0.04$ & $1.37\pm0.01$ & $1.52\pm0.01$\\
\hline
&& $L_{\rm cb}=30\mpc$& \\
 \hline
blue  & $0.84\pm0.08$ & $0.99\pm0.02$ & $1.16\pm0.03$\\
red  & $1.37\pm0.15$ & $1.35\pm0.02$ & $1.53\pm0.02$\\
\hline
\end{tabular}
\end{center}
\end{table}

\section{Cosmic web in the SDSS volume}
\label{sec_web}

\begin{figure}
\centering
\includegraphics[width=.5\textwidth]{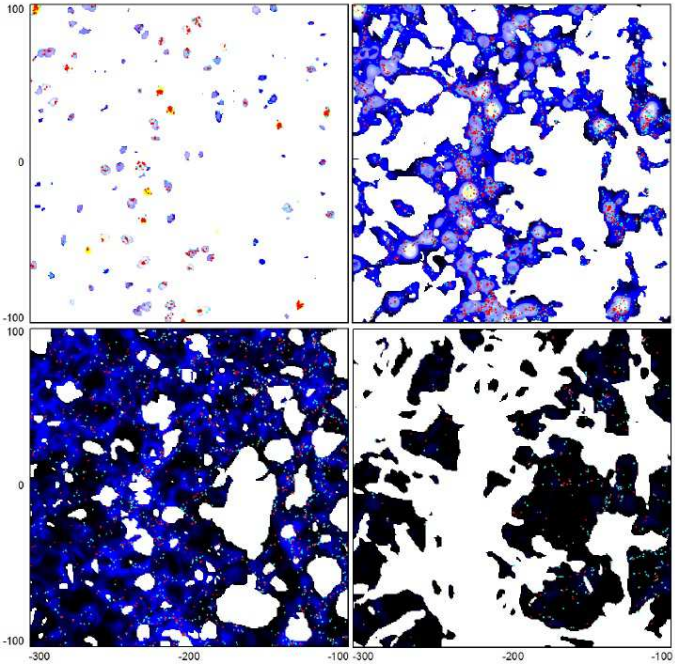}
\caption{The projected density fields and galaxies in volumes classified as knots (top left), filaments(top right),
sheets (bottom left) and voids (bottom right). The red and cyan points represent red and blue galaxies respectively.
The color code for the density field and the slices are the same as those in Figure \ref{fig_dggh}.}
\label{fig_lssgal}
\end{figure}

One of the most prominent structures in our local Universe is the Sloan
Great Wall which is at $z\sim0.08$. Figure \ref{fig_dggh} shows part of this massive
structure, which contains several galaxy clusters.  A large, low-density region
containing only a few galaxies lies on the right side of the Great Wall.
Galaxy formation efficiency appears to be very different in the two different regions.
In order to understand this, it is useful to classify the large scale structures
into different categories to represent their dynamical properties. As discussed in
Section \ref{sec_sts}, local tidal tensor may provide a useful way to classify
the cosmic web. Here we show the classification using $R_{\rm s}=2\mpc$
and $\lambda_{\rm th}=0.2$ (see e.g. Forero-Romero et al. 2009).
The projected density maps for the four different web types, knots, filaments,
sheets and voids, are separately shown in the four panels of Figure \ref{fig_lssgal}.
The slice is the same as that shown in Figure \ref{fig_dggh}, i.e. of $10\mpc$ thick.
At any position on the projected 2-dimensional map, there are 10 grid cells
projected along the direction perpendicular to the slice. Since it is
likely that the 10 overlapping grids belong to different categories,
there are overlaps between different web types.
If no grid cell at a given position in the slice is classified as the web type in question,
the position at the corresponding panel is plotted in white color. One can see
that the Sloan great wall is identified as a filamentary structure, percolating
the whole slice in the vertical direction, while the low density region is classified
as a void. Cells classified as sheet occupy a very large part of the plot,
although the volume of the sheet cells is only about 40\% of the total.
This implies that the sheet structures are thin.
Galaxies are also separated into four types, based on the locations of their host
groups, and their positions are superposed on the corresponding projected
density fields.
The knot galaxies have the highest galaxy density and the FOG effect has been corrected by our method.
The distribution of filament galaxies is more concentrated than those of sheet and void galaxies.
Inspecting the two low-density environments (sheets and voids), we see that
galaxies tend to reside in relatively high density regions. In particular,
red galaxies are more abundant than blue ones in knots but less abundant in voids,
in broad agreement with previous results based on galaxy clustering (e.g. Li et al. 2006).

\begin{figure}
\centering
\includegraphics[width=0.5\textwidth]{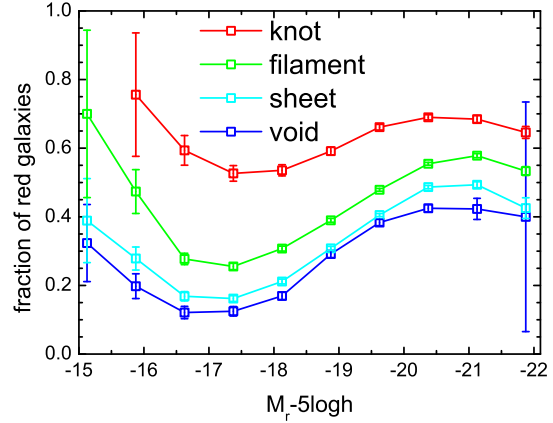}
\caption{The fractions of red galaxies as functions of the $r$-band absolute magnitude.
The color-coded lines show the results for galaxies residing in knots, filaments, sheets and voids as indicated.}
\label{fig_redf}
\end{figure}

To investigate the environmental dependence of galaxy color
quantitatively, we show in Figure \ref{fig_redf} the fraction of red galaxies in the
four environments as a function of the $r$-band magnitude.  For all the luminosity
bins, the red fraction is the highest for knot galaxies, followed successively
by filament, sheet and void.  For knot galaxies, the fraction is slightly more
than 60\% on average and exhibit a weak dependence on galaxy luminosity
if the data point at the lowest luminosity, which has a large error bar, is excluded.
For filament, sheet and void galaxies, the fraction varies with luminosity in a
non-monotonous way, in the sense that it first decreases then increases as
the luminosity increases. These results reflect the complicated quenching processes, such as environmental quenching and mass quenching (e.g. Peng et al. 2010). Since we know that the halo population is different in
different structures (see Hahn et al. 2007), one interesting question is
whether these environmental processes can be explained in terms of
the differences in the halo population.  Studying this problem may help us to
understand the role of large scale structure in galaxy formation. In a future work,
we will use the CS and group catalog to investigate this question further.

\begin{figure}
\centering
\includegraphics[width=0.5\textwidth]{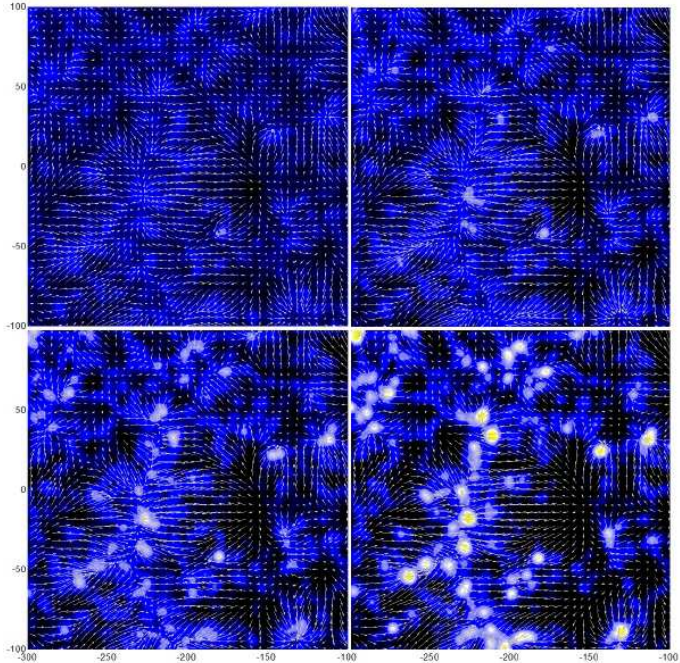}
\caption{The evolution of the density (color maps) and velocity (arrows) fields in a slice of 10$\mpc$ thick
enclosing part of the Sloan Great Wall.
The four panels show the results at $z=4.04$ (top-left), $z=2.03$(top-right), $z=0.99$ (bottom-left) and $z=0$ (bottom-right),
respectively. The color coding is the same as that in Figure \ref{fig_drec}. The length of a arrow is
proportional to the magnitude of the velocity it represents.}
\label{fig_evogw}
\end{figure}

One advantage of the constrained simulation is that it can reveal the dynamical
state and formation history of the large scale structures. In Figure \ref{fig_evogw} we
show the density fields together with the velocity fields around the Sloan great wall
at four different redshifts. If searching for the Sloan great wall only through the density
contrast, one cannot see any sign for the great wall before $z\simeq4$, but it
starts to show sign at $z\simeq2$. By $z\simeq1$ the great wall grows
into a dominant structure in the slice. Since then it becomes more concentrated
but its co-moving length changes only little. Its strong gravity empties the nearby
region, forming a big void at the right. On the other hand, the evidence for the
existence of the wall  can be seen in the velocity field  at $z\simeq4$.
The velocity field clearly reveals that the great wall continues to attract material
from the left and right sides during almost the entire history of the Universe. Compared
to the perpendicular component, the velocity component along the filament is
very small, so that the collapse along this direction is prolonged.  This
is consistent with the theoretical expectation for the mass flows around
filamentary structures, and such flow pattern can affect the dynamical properties
of the halos residing in filaments (Shi et al. 2015).

\begin{figure}
\centering
\includegraphics[width=0.5\textwidth]{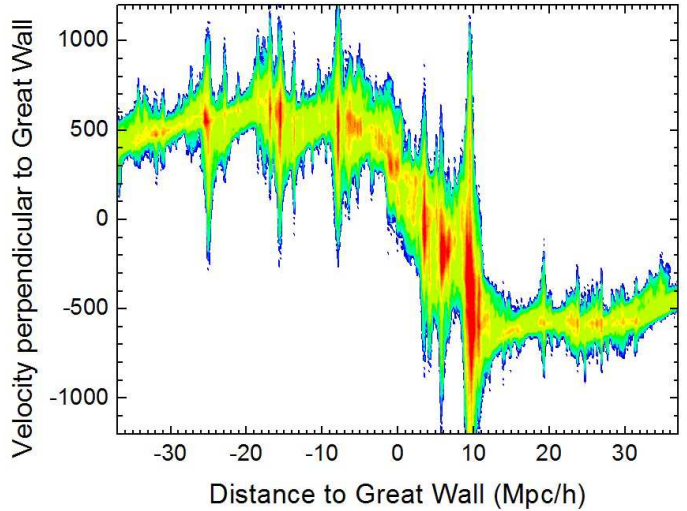}
\caption{The phase-space distribution for dark matter particles within a column. The symmetric axis of the column is
perpendicular to the Sloan great wall and its radius is $5\mpc$. The velocities and distances shown are all
projected onto the symmetric axis of the column. Distance equal to zero indicates the great wall location.
Red color denotes high phase-space density while blue denotes low phase-space density. White regions are empty
of particles.}
\label{fig_gwv}
\end{figure}

To show the accretion flow around the Sloan great wall in more details, we define a column that
is perpendicular to the filament and show the phase-space distribution for dark matter particles
within this column in Figure \ref{fig_gwv}. To avoid the contamination of particles in massive clusters,
we place the center of the column between two massive clusters, which are separated by
about 19$\mpc$. The radius and length of the column are set to be $5\mpc$ and $80\mpc$,
respectively. The velocity and distance (to the center of the column) are both projected onto the
rotation axis of the column. One can clearly see that the materials in the column flow towards
the filament from both the near and far sides. In the outer region where the distance to the
great wall is larger than about 5 $\mpc$, the mean velocity remains more or less
at a constant level, around $500\kms$. Since the great wall is quite straight
(Figure \ref{fig_drec}) and much longer than the scale considered here, its gravity
over this scale depends only weakly on the distance to it. In the linear regime, the velocity
is proportional to the gravity, and so the constant velocity may be explained
by the straight filamentary structure. In the inner region, on the other hand, the flow velocity
decreases with decreasing distance, indicating that the width of the great way is less than
$10\mpc$. The velocity dispersion in the inner region appears larger than that in the outer
region -- note again that the column is not centered on any cluster,  which is likely caused
by the collision of the convergent flow. Such collisions are expected to heat the
associated inter-galactic medium to high temperatures, which may be observed
through UV absorption, X-ray emission, and Sunyaev-Zel'dovich effect.
Moreover, the heating might also quench gas accretion into small halos embedded
in the filaments, eventually affecting galaxy evolution in such environments
(Mo et al. 2005). The Sloan Great Wall is therefore a wonderful place for studying the
inter-galactic medium and the interaction between galaxies and the large-scale
structure. Constrained hydrodynamical simulations are needed in order to address
this problem in detail.

\begin{figure}
\centering
\includegraphics[width=0.5\textwidth]{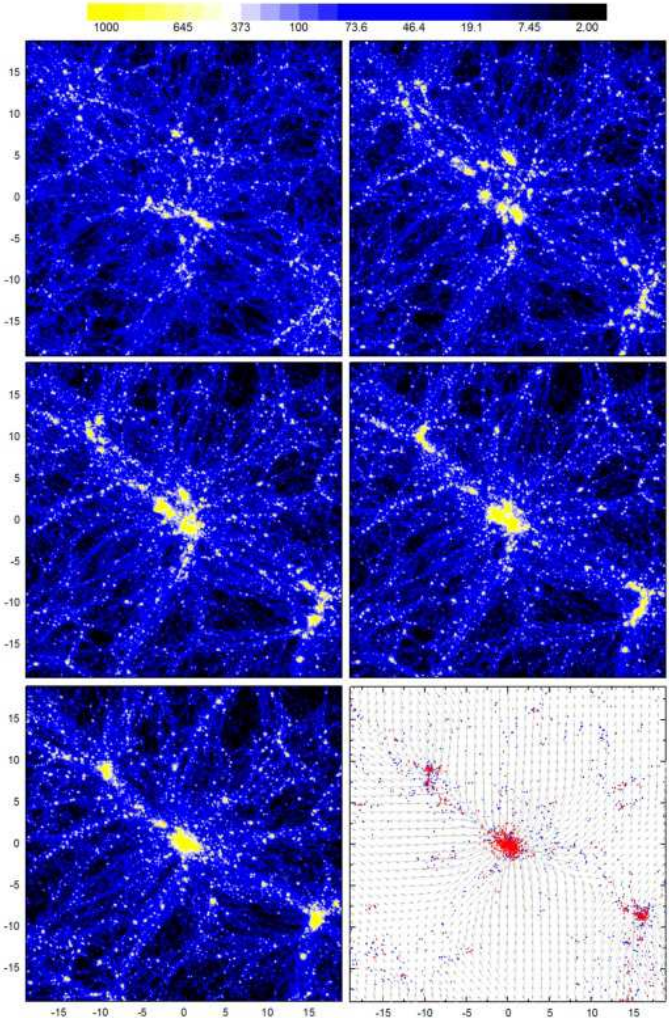}
\caption{The projected mass density fields in a slice of $39\times39\times10\mpc$ centered on the
Coma cluster at $z\sim2$ (top-left), $z\sim1$ (top-right), $z\sim0.5$ (middle-left), $z\sim0.3$ (middle-right) and
$z=0$(bottom-left). The bottom-right panel shows the spatial distribution of real galaxies and the velocity field
extracted from the CS.}
\label{fig_coma}
\end{figure}

\begin{figure}
\centering
\includegraphics[width=0.5\textwidth]{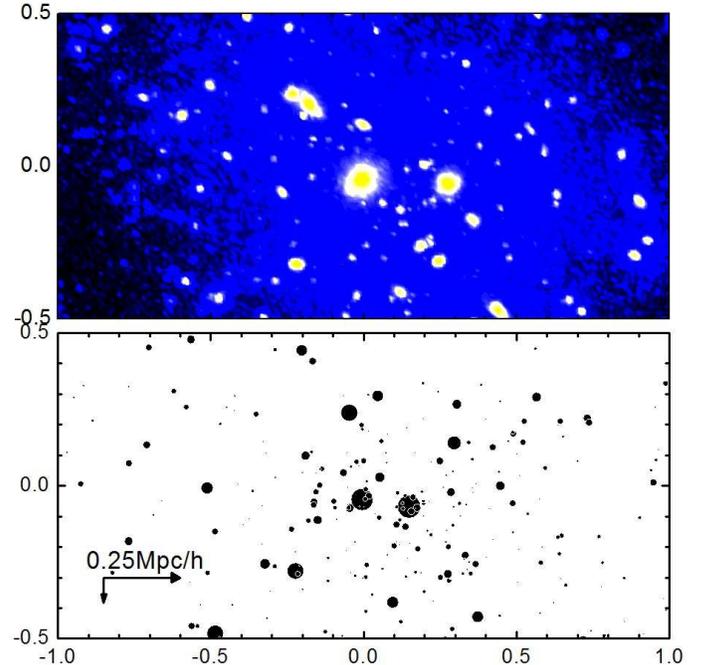}
\caption{The upper panel shows the inner region of the Coma cluster in the CS. Different from other
density maps, this figure is color-coded by the square of the projected densities so that the substructures
can be seen more clearly. In lower panel, the black dots show the spatial distributions of real galaxies in
the Coma galaxy cluster. The symbol sizes are scaled with the galaxy luminosities, and all galaxies are
shifted horizontally by $\sim0.25\mpc$ and vertically by $\sim0.08\mpc$ (as indicated by the arrows).}
\label{fig_comasub}
\end{figure}

Another interesting object in the nearby Universe is the Coma galaxy cluster.
The simulated halo corresponding to this cluster has almost the same mass and
position: the Coma galaxy cluster has an estimated mass of
$9.1\times10^{14}\msun$ in Yang et al (2007) catalog, and the corresponding halo is
only 0.04 dex larger, about $1.0\times10^{15}\msun$;
the spatial separation between the real cluster and the simulated halo is only
$0.018\mpc$. In addition, they also have similar shapes in projection.
Figure \ref{fig_coma} shows the projected CS density map and galaxy distribution in a
slice of $39\times39\times10\mpc$ centered on the Coma cluster. The projected direction is chosen to be parallel with the
line of sight to the Coma cluster. One can see that the galaxy distribution within
the Coma cluster aligns with the mass distribution of the halo. It is known that massive
halos are strongly aligned with the large scale structures within which they are
embedded (e.g. Hahn et al. 2007; Wang et al. 2011). Since our method can
reliably recover the large scale structures, it is not surprising to find the
similar alignments between the observed galaxy distribution and the CS
mass distribution. Nevertheless, it suggests that the CS can be used
to study various alignments between galaxies/galaxy systems with the large
scale structures (e.g. Brown et al. 2002; Faltenbacher et al. 2009; Zhang et al. 2013).

Figure \ref{fig_coma} illustrates again how our method can reliably recover the
large scale structures. For instance,  in addition to the Coma cluster, the other two
massive groups, located respectively in the bottom-right and top-left corners,
also have corresponding massive halos. The massive filament connecting the
three galaxy systems and the surrounding big voids are also reproduced by the CS.
The velocity field obtained from the CS is superposed on the galaxy distribution
in the bottom-right panel. The flows converge to the three massive galaxy
groups/clusters. A filamentary flow pattern towards the Coma cluster is also clearly
seen,  which corresponds to the massive filament in the density map.
Recall that the flow along the Sloan great wall is quite slow, which is very
different from what we are seeing here.

The CS reveals the possible formation history of the Coma cluster and the nearby structure
(Figure \ref{fig_coma}). At $z\simeq2$, only several compact and thin filamentary structures
embedded in an over-density region are produced.  These filaments collapse and merge
to form a number of halos at $z\sim1$. These halos then move close to each other and merge
to form two massive halos of comparable sizes surrounded by numerous small halos.
Meanwhile, thin filaments, which appear parallel, are assembling into thick filaments
connected to the Coma cluster observed at $z=0$. The velocity field reveals that the
Coma cluster is still accreting material through the filaments even at the present day.
The CS suggests that the Coma cluster has experienced a violent merger in the
recent past ($z\sim 0.3$). The upper panel of Figure \ref{fig_comasub} shows the zoom-in of
the dark matter halo of the Coma cluster at $z=0$.
Two high density peaks in the (projected) central region, which are the vestiges of the
major merger, are clearly seen.  It is interesting to compare these results with the
observational properties of the Coma cluster.  In the lower panel of Figure \ref{fig_comasub}, we
show the galaxy distribution in the central region of the Coma cluster with a slight
($\sim0.25\mpc$) offset. The spatial distribution of luminous galaxies resembles
that of the massive substructures. In particular,  the two supergiant elliptical galaxies in the central
region (see also Fitchett \& Webster 1987) roughly match the two high density peaks,
and may indeed correspond to the central galaxies that formed earlier in the two massive
progenitors seen at $z>0.3$. Observations at other wave bands also support this merger
scenario for Coma.  For example, X-ray observations suggested that the Coma cluster is
unrelaxed (e.g. Briel  et al. 1992; Sanders et al. 2013); radio observations revealed the presence
of a radio relic (Giovannini et al. 1985), which might be produced by shock waves
(En{\ss}lin et al. 1998; Miniati et l. 2001) produced by the merger of the two massive progenitors.
These results clearly demonstrate the power of our reconstruction in understanding the
histories and structures of massive objects observed in the nearby Universe.
A movie, showing the 3-dimensional structure of the simulated Coma cluster and
the comparisons with optical and X-ray observations, is shown at
\url{http://staff.ustc.edu.cn/~whywang/MOVIE/}.

\begin{figure}
\centering
\includegraphics[width=0.5\textwidth]{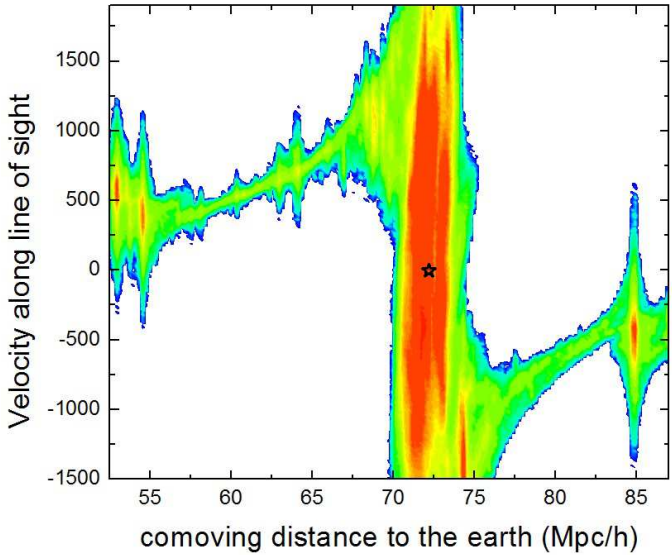}
\caption{The phase-space distribution for dark matter particles within a column. The symmetric axis of the column
is the line of sight to the Coma cluster and its radius is $2.5 \mpc$.  The velocities and positions shown are
projected onto the line of sight. The star indicates the velocity and position of the Coma cluster.
Red color denotes high phase-space density while blue denotes low phase-space density.
White regions are free of particles.}
\label{fig_comavel}
\end{figure}

Finally, in Figure \ref{fig_comavel}, we show the phase-space distribution of dark matter
particles within a column, with its axis along the line of sight (LOS) to the Coma cluster
and a radius of $2.5\mpc$, comparable to the virial size of the Coma cluster.
The velocity and position are both projected onto the LOS, with the cluster center
denoted by a star.  Due to the gravitational attraction of the cluster, the bulk
velocity of the surrounding material increases with decreasing distance to the cluster,
reaches as high as $1000\kms$ on average.  At locations far from the cluster,
dark matter particles occupy a very narrow range in velocity, indicating that
the velocity dispersion of the structure is quite low. However, as the distance
to the Coma decreases to about $5\mpc$ (about 2 times of the virial radius),
the velocity dispersion starts to increase with decreasing distance,   and
the increase becomes very rapid as the distance approaches $\sim2\mpc$.
Within $2\mpc$, the phase-space distribution are dominated by random motion
within the cluster while the bulk motion becomes negligible (see the black star).
The sharp edge indicates that the cluster size is about 2$\mpc$.
The presence of substructures within $2\mpc$ can be seen clearly,
consistent with the results shown above.

The examples described above demonstrate clearly that the constrained simulations
can provide much environmental and dynamical information for the observed galaxies
and galaxy systems. To check the reliability of such environmental and dynamical
information, we have made tests using mock catalogs and estimate the uncertainties
of the predicted tidal and velocity fields. The results are presented in the appendix.
To summarize, the uncertainties in the three eigenvalues of the tidal
tensor estimated from the CS density field are similar, about $0.2$ for
a smoothing radius $R_{\rm s}=4\mpc$, and increasing to $\sim 0.6$ at
$R_{\rm s}=2\mpc$. The uncertainty in the velocity field depends on the distance
to the boundary of the reconstruction volume, because the structures outside
the reconstruction volume are not constrained. We divide the reconstruction volume into
three parts, the inner, intermediate and boarder regions, according to their ``distances'' to
the boundary. The uncertainties of the one dimensional velocity in the three regions
are about 80, 132 and $195\kms$, respectively, at the smoothing radius of $4\mpc$.
As the smoothing scale decreases to $2\mpc$, the uncertainties remain more or
less the same. This is expected because gravitationally generated
velocities are dominated by large scale modes of the cosmic density field.

\section{Summary}
\label{sec_sum}

In our previous studies (Yang et al. 2005;2007; Wang et al. 2009;2012;2013; Paper I) we have developed a
series of methods to reconstruct the initial density field from a galaxy redshift survey.
These methods have been tested in great detail with $N$-body simulations and
mock galaxy catalogs.  In this paper, we apply these methods to the SDSS DR7 galaxy
sample at $z\leq0.12$. We start with galaxy groups identified from this sample
by the halo-based group finder of Yang et al. (2005, 2007) and use them to represent
dark matter halo population. Redshift distortions in the distances of
the groups are corrected by using the peculiar velocities of the groups predicted
by the reconstructed density field. We then adopt the halo-domain method of Wang et al. (2009)
to reconstruct the present-day density field from the distribution of the halo population
in real space. The initial density field for the formation of the  local universe is
reconstructed from the present-day density field by using the Hamiltonian Markov Chain Monte
Carlo algorithm combined with the PM dynamics that evolves the initial density field to
the density field today (Paper I).  And finally, the reconstructed initial density field is used
as the initial condition for an $N$-body simulation to obtain a constrained simulation (CS).
This CS assumes WMAP5 cosmology and trace the evolution of $3072^3$ particles in a
periodic cubic box of $500\mpc$ on a side. Tests with mock galaxy catalog indicate that
the reconstruction is reliable, with an uncertainty in the mass density of about $0.10$ ($0.23$)
dex, in the tidal field of about $0.2$ ($0.6$), in the velocity field of
$\leq130\kms$ ($\leq136\kms$)  on a scale of $4\mpc$ ($2\mpc$) at $z=0$. Our investigations
show that the main sources of the uncertainties in the reconstruction are the redshift
distortion and the inaccuracy of the group finder.

We present results for a number of important statistics of the constrained simulation
on various scales, such as the mass power spectra, the volume fractions of different types
of the cosmic web, the halo mass function, and their evolutions with time.
We also identify sub-halos and construct merger trees for both halos and sub-halos.
The results are in good agreement with theoretical predictions and with those given by a reference simulation
that assumes random initial phases. All these demonstrate that our reconstruction is unbiased.

We then investigate the relation of the CS with real groups and galaxies. We cross match the observed
groups with the FOF halos identified from the CS. About half of the SDSS groups with masses larger
than $10^{14}\msun$ have matched halos with distances less than $1\mpc$ and mass differences
less than 0.15 dex. Choosing the maximum distance difference to be $4\mpc$ and the maximum mass
difference to be 0.5 dex results in 95\% and 80\% of the groups with $M_{\rm gr}\geq10^{14}\msun$
and $\geq10^{13}\msun$ having matched halo. The matched fractions are much higher than
those between the groups and halos in a rotated sample, which are $<$5\% and 10\%
for the above two cases, respectively.  For smaller groups ($\geq10^{12}\msun$), the matched fraction is 85\%,
slightly larger than  that between the groups and rotated halo sample, which is 40\%. Thus, the one-to-one match
is robust for cluster-sized halos but marginally so for galactic-sized halos. However,
the number density of groups with $10^{12}\msun\leq M_{\rm gr}\leq10^{13}\msun$
in small cubes (20$\mpc$ on a side) is tightly correlated with that of CS halos in the
same mass range, much tighter than that between random pairs of cubes.
This suggests that the statistics for small groups are also well reproduced in volumes much
smaller than the reconstruction volume, and so cosmic variance is effectively reduced in the
constrained simulation. The reliability of the matching depends  directly on the accuracy of the PM
model used in the HMC. One can, therefore,  improve the reconstruction by adopting PM models
of higher accuracy. The good match between massive groups and CS halos suggests that
one can rely on the CS to study the histories of individual galaxy systems, at least for massive ones.
We use the Coma cluster and the SDSS great way as examples to demonstrate how the CS
can be used to study the formation history, density and velocity field in and around known observed
structures.

We also investigate the correlations between the mass density obtained from the CS
and galaxies in the reconstruction volume. Strong correlations are found, and bias in the distribution
of galaxies with respect to the mass is clearly seen, being stronger for high-luminosity and red galaxies.
The correlation with the mass density field is found to be tighter for red galaxies than for blue
galaxies of the same luminosity, indicating that red galaxies are better in tracing the mass density.
This connection between the observed galaxies and the underlying mass density  provides a direct
way to investigate how the bias depends on galaxy properties and how the scatter between the
two density fields is produced. We will come back to a detailed investigation on this in a future paper.
Based on the CS density field, we classify the cosmic web in the SDSS volume into four categories,
knot, filament, sheet and void, based on the local tidal tensors obtained from the CS.
We find that the fraction of red galaxies is the largest in knots, followed by filaments, sheets
and voids, for almost all luminosity ranges considered.

The constrained simulation presented in this paper have many other applications in the future.
For instance, we can model galaxy formation using halos and merger trees extracted from the CS
combined with the galaxy formation models, such as the semi-analytic model and other
halo-based empirical models. The modeled galaxies will guaranteed to have similar spatial distribution
to the real galaxies, and environmental effects in halo formation will be automatically taken into
account. This will make the comparison between models and observations statistically
more robust and free of cosmic variance.  Furthermore, we can carry out gas simulations directly from
the reconstructed initial condition to study the interactions between galaxies, gas, and dark
matter in different regions of the observed universe.   Such an approach is particularly useful in probing the gas
components and its evolution in massive large scale structures and massive galaxy clusters, for which
there are one-to-one correspondences between the observation and the CS.
For example, the converging mass flow towards the Sloan great wall has quite high velocities.
The gas simulation can help us to understand the state of the gas associated with the great wall,
and to examine the influence of such a large structure on the cooling and condensation
of gas in dark matter halos embedded in the structure.  The constrained simulations can also provide
information about the large-scale environments and their dynamical states for galaxies and galaxy
systems. An analysis of the correlation between galaxy intrinsic and environmental properties
can then be made straightforwardly, which can provide constraints on how galaxies form and evolve
in the cosmic density field.

\section*{Acknowledgments}

This work is supported by the 973 Program (2015CB857002,2015CB857005), NSFC
(11522324,11421303,11233005,11533006,11320101002),  the Strategic Priority Research Program ``The
Emergence of Cosmological Structures" of the Chinese
Academy of Sciences, Grant No. XDB09010400 and the Fundamental
Research Funds for the Central Universities. H.J.M.
would like to acknowledge the support of NSF AST-1517528.
The work is also supported by the Supercomputing Center
of University of Science and Technology of China and the Center for High
Performance Computing, Shanghai Jiao Tong University.

\appendix

\section{Tests with mock catalog}\label{app_mk}

To estimate the uncertainties of our reconstruction, we use a mock galaxy catalog to
test our methods described in Section \ref{sec_method}.
The mock galaxy catalog is exactly the same as that used in Wang et al. (2012; 2013) and constructed
in a way similar to that described in Yang et al. (2007).  The mock galaxy catalog is built from the
`Millennium Simulation' (MS) carried out by the Virgo Consortium (Springel et al. 2005). We populate galaxies
in dark matter halos identified from the MS according to the conditional luminosity function model
(CLF)  of van den Bosch et al. (2007) and assign phase-space coordinates (positions and peculiar velocities)
to these galaxies following Yang et al. (2004). Note that the peculiar velocities of galaxies include the contribution
of both the motions of the host halos and virial (random) motion of galaxies within their host halos.
We then stack $3\times3\times3$ galaxy-populated simulation boxes together and place
a virtual observer at the center of the stack. With respect to the observer, each galaxy is
assigned ($\alpha_{\rm J}$, $\delta_{\rm J}$) coordinates and a redshift, which is a combination
of its cosmological redshift and peculiar velocity. Finally, a mock galaxy catalog is constructed by
mimicking the sky coverage of the SDSS DR7, taking into account the angular variation of
the magnitude limit and survey completeness (see Li et al. 2007).  Readers are referred  to
Wang et al. (2012; 2013) for details.

We apply the halo-based group finder of Yang et al. (2005; 2007) to the mock galaxy catalog
and  obtain a mock catalog of groups, with masses assigned according to the characteristic
luminosities of individual groups (Section \ref{sec_gr}). Redshift distortions are corrected  for the
mocked groups and the halo-domain method is used to reconstruct the present-day density field ($\rho_{\rm hd}$),
following exactly the same methods described in Section \ref{sec_crd} and \ref{sec_hd}. We then
reconstruct the initial density field  from $\rho_{\rm hd}$, using the HMC$+$PM method
with the same spatial and time resolutions as those described  in Section \ref{sec_ic}.
Finally, the reconstructed initial density field is  evolved to the present day with Gadget-2.
Thus, all procedures used here for the mock catalog are exactly the same as that given in Section
\ref{sec_method} for the SDSS catalog. The only main difference is that here we use the
same cosmology as adopted in MS, which is different from WMAP5 adopted in this paper,
and a lower mass resolution for the constrained simulation ($800^3$ particles).

\begin{figure}
\centering
\includegraphics[width=0.7\textwidth]{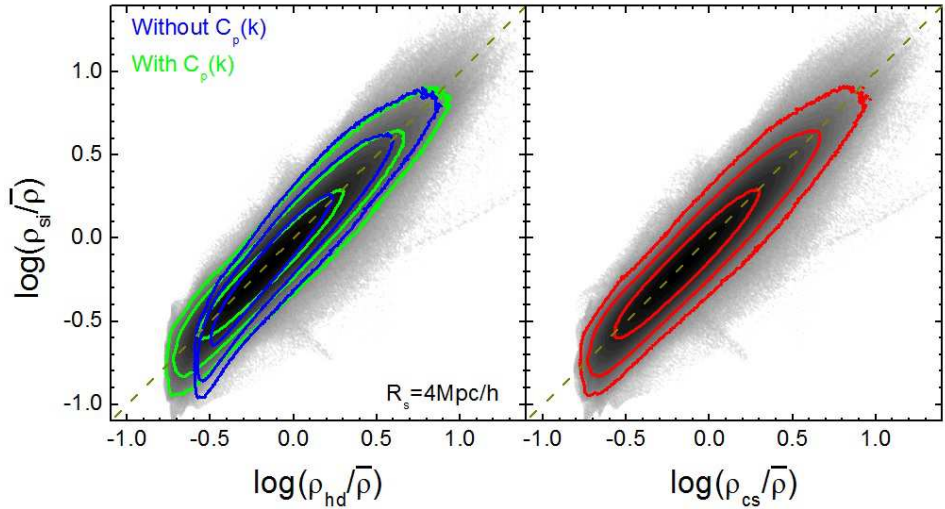}
\caption{Density-density plots between the original MS simulation($\rho_{\rm si}$) and the
halo-domain reconstruction ($\rho_{\rm hd}$, left panel), and between the MS and the CS ($\rho_{\rm cs}$, right panel).
The density fields are smoothed with a Gaussian of radius $R_{\rm s}=4\mpc$. The three contours encompass
67\%, 95\%, and 99\% of the grid cells in the reconstruction volume. The dashed lines indicate the perfect
relation. In the left panel we also include the results for the halo-domain reconstruction that does not include
the correcting function ($C_{\rm p}(k)$) for comparison (blue contour). Please see Appendix for details.}
\label{fig_mockd}
\end{figure}

The density-density relation between $\rho_{\rm hd}$ and the original density field
$\rho_{\rm si}$, both smoothed with $R_{\rm s}=4\mpc$, is shown in the left panel of Figure \ref{fig_mockd}.
One sees that $\rho_{\rm hd}$ follows $\rho_{\rm si}$ without significant bias. The typical scatter, in terms of the r.m.s of
$\log{(\rho_{\rm si}/\rho_{\rm hd})}$,  is about 0.09 dex. For a smaller smoothing radius, $R_{\rm s}=2\mpc$,
the scatter increases to 0.20 dex. For comparison, we also show the density field from the halo-domain
reconstruction without applying any correction to the Fourier modes (see Section \ref{sec_hd} for details).
As one can see, the correction function lowers the density at the lowest density end of
$\rho_{\rm hd}$ while having little effects at in high density regions, making $\rho_{\rm hd}$
better in matching the original density field, $\rho_{\rm si}$. The right panel of Figure \ref{fig_mockd}
presents the comparison between the density fields of the constrained simulation (CS) and the
original simulation. The correlation is very similar to that between $\rho_{\rm hd}$ and
$\rho_{\rm si}$. This is expected as the correlation between $\rho_{\rm hd}$ and
$\rho_{\rm cs}$ is very tight, with scatter of only $0.02$ dex.
Overall, the CS density field matches the original one very well;
the typical scatter between the two is 0.10 dex at $R_{\rm s}=4\mpc$ and 0.23 dex at
$R_{\rm s}=2\mpc$. These tests demonstrate that our HMC$+$PM method is accurate;
the main errors are produced by the reconstruction of the current density field, i.e.
from the uncertainties in redshift distortion correction, in the group finder, and in the
density model based on halo domains.

\begin{figure}
\centering
\includegraphics[width=0.7\textwidth]{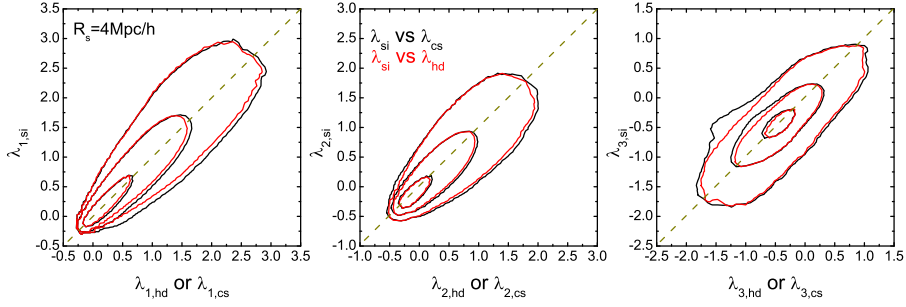}
\caption{The three eigenvalues of the tidal tensor obtained from the original simulation ($\lambda_{\rm si}$) versus
those obtained from the halo-domain reconstruction ($\lambda_{\rm hd}$, red) and the CS ($\lambda_{\rm cs}$, black).
The three contours encompass 67\%, 95\%, and 99\% of the grid cells in the reconstruction volume.
The dashed lines represent the perfect relation. The fields are smoothed on a scale of $4\mpc$.}
\label{fig_mockt}
\end{figure}

Next we consider the cosmic tidal field,  which provides important environmental information for
studying galaxies and dynamical information to study the large-scale structure.
Figure \ref{fig_mockt} shows the three eigenvalues of the tidal tensor calculated from the CS
($\lambda_{i,\rm cs}$) versus the corresponding quantities calculated directly from the MS simulation
($\lambda_{i,\rm si}$). The tidal fields are calculated using Eqs. (\ref{eq_tij}) and
(\ref{eq_phi}), and are smoothed on the scale of $4\mpc$. Overall, $\lambda_{i,\rm cs}$ are strongly correlated
with $\lambda_{i,\rm si}$, without significant bias. The typical scatter in terms
of the r.m.s in $\lambda_{i,\rm si}-\lambda_{i,\rm cs}$,  is around 0.2 (0.21 for the major, 0.20 for the
intermediate, and 0.20 for minor axes, respectively). Adopting a smaller smoothing scale of $2\mpc$,
increases the scatter to about 0.6.  This increase is almost certainly due to the fact that small
scale structures have significant contribution to the tidal field, but are more difficult to be
constrained accurately than large scale structures.

For comparison, we also show the results for the tidal field calculated based on
$\rho_{\rm hd}$. The corresponding eigenvalues $\lambda_{i,\rm hd}$ are slightly more tightly correlated
with the original ones, with typical scatter of 0.20, 0.18 and 0.19 for the three axes,
respectively. This shows again that the uncertainties produced by the HMC$+$PM is small.

\begin{table}
\begin{center}
\caption{Uncertainties in the reconstructed density, tidal and velocity fields at $z=0$}
\label{tab_err}
\begin{tabular}{lcccc}
\hline\hline
${\rm density}$ & $\rho_{\rm hd}$ & $\rho_{\rm cs}$\\
\hline
$R_{\rm s}=4\mpc$  & 0.09 $\rm dex$ & 0.10 $\rm dex$\\
$R_{\rm s}=2\mpc$  & 0.20 $\rm dex$ & 0.23 $\rm dex$\\
\hline
\end{tabular}
\end{center}
\begin{center}
\begin{tabular}{lccccccccccccccccccccc}
\hline
\hline
${\rm tidal}$ & $\lambda_{1,\rm hd}$ & $\lambda_{2,\rm hd}$ & $\lambda_{3,\rm hd}$ & $\lambda_{1,\rm cs}$ & $\lambda_{2,\rm cs}$ & $\lambda_{3,\rm cs}$\\
\hline
$R_{\rm s}=4\mpc$  & 0.20 & 0.18 & 0.19 & 0.21 & 0.20 & 0.21 \\
$R_{\rm s}=2\mpc$  & 0.60 & 0.55 & 0.53 & 0.63 & 0.59 & 0.57\\
\hline
\end{tabular}
\end{center}
\begin{center}
\begin{tabular}{lccccccccccccccccccccc}
\hline
\hline
${\rm velocity}$ & $v_{\rm cs}($\rm inner$)$ & $v_{\rm cs}($\rm middle$)$ & $v_{\rm cs}($\rm boundary$)$ \\
 \hline
$R_{\rm s}=4\mpc$  & 79.9$\kms$ & 131.9$\kms$ & 194.7$\kms$\\
$R_{\rm s}=2\mpc$  & 86.5$\kms$ & 135.7$\kms$ & 198.5$\kms$\\
\hline
\end{tabular}
\end{center}
\end{table}

\begin{figure}
\centering
\includegraphics[width=0.7\textwidth]{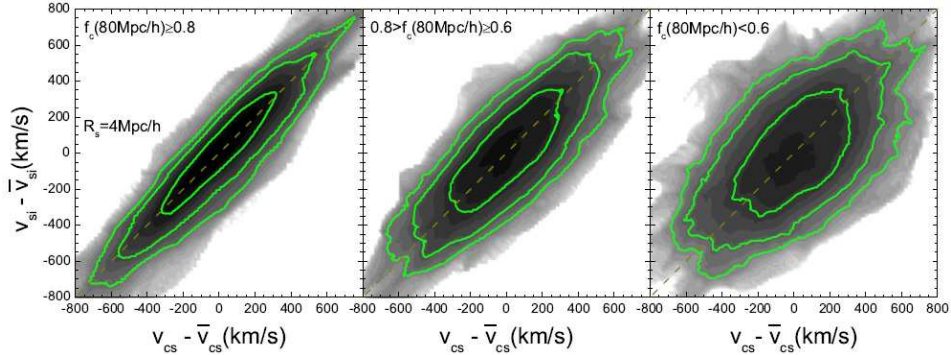}
\caption{The one dimensional velocity obtained from the CS versus the velocity in the original simulation.
Results are shown for grid cells with three different $f_{\rm c}(80\mpc)$, which is used to indicate the closeness to the
boundary (see text for definition). The dashed lines indicate a perfect relationship.
All the velocity fields are smoothed on a scale of $4\mpc$.}
\label{fig_mockv}
\end{figure}

Finally, we compare the velocity fields derived from the CS ($v_{\rm cs}$) with the true velocity in the original simulation
($v_{\rm si}$). Because the structures outside the reconstruction volume are unconstrained, the motion induced by these
structures cannot be properly recovered by our methods.
The uncertainty caused by such effect is expected to be stronger
near the boundary of the reconstruction volume.  To quantify this effect, we define a parameter to indicate
the distance to the boundary. Following Wang et al. (2012, see also Section \ref{sec_ic}), we use $f_{\rm c}(R_{\rm b})$
with $R_{\rm b}=80\mpc$ to indicate the closeness of a grid to the boundary. For a grid cell close to the boundary
the value of $f_{\rm c}(80\mpc)$ is expected to be smaller than one, while for a grid cell located more than
$R80\mpc$ away from the survey boundary, $f_{\rm c}(80\mpc)=1$. We divide the reconstruction volume into three
parts according to the values of $f_{\rm c}(80\mpc)$ and examine them separately.
Furthermore, since the mass fluctuations on scales larger than our reconstruction volume may
induce a bulk motion in the whole reconstruction volume, we also subtract the mean velocity
in the whole reconstruction volume from all velocities used in our calculation.
Figure \ref{fig_mockv} shows $v_{\rm cs}-\bar{v}_{\rm cs}$ versus $v_{\rm si}-\bar{v}_{\rm si}$ for
grid cells in the inner region ($f_{\rm c}(80\mpc)\geq 0.8$), the intermediate region
($0.8>f_{\rm c}(80\mpc)\geq 0.6$) and the boundary region ($f_{\rm c}(80\mpc)<0.6$).
Here, $v_{\rm cs}$ and $v_{\rm si}$ represent a single velocity component along one of the three axes of
the Cartesian coordinates,  and $\bar{v}_{\rm cs}$ and $\bar{v}_{\rm si}$ are the
corresponding mean velocities in each of the $f_c$ regions (their absolute values are less than $50\kms$).
Since the results for the three velocity components are similar, we plot them together.
The velocity fields shown in the figure are smoothed with $R_{\rm s}=4\mpc$.
In the inner region, the reconstructed velocity is linearly correlated with the true velocity, with slope
very close to unity. The typical scatter, in terms of r.m.s in $v_{\rm si}-v_{\rm cs}$,  is about 80$\kms$.
As $f_{\rm c}$ decreases, the slope remains around unity,  but the scatter increases.
For grid cells in the intermediate and boundary regions, the values of the scatter are
$132$ and $195\kms$, respectively. When smoothed with a smaller $R_{\rm s}$ of $2\mpc$,
the scatter increases only slightly to $87$, $136$ and $199\kms$ for the three $f_c$ regions.
Note that the volumes of the inner, intermediate and boundary regions are 45\%, 31\% and 24\% of the
total reconstruction volume, respectively.

We have also checked the dependence of the uncertainties in the density and
tidal fields on $f_{\rm c}$. The uncertainties for the two quantities are found to be almost independent
of the distance to the boundary so we do not show the results. Note that all of these uncertainties are
estimated at $z=0$, and listed in Table \ref{tab_err}.

\end{document}